\documentclass[twocolumn]{aastex631}
\usepackage{rotating}

\accepted{20 November 2023}
\submitjournal{AAS Journals}

\shorttitle{Formation of $\beta$ Pic b}
\shortauthors{Reggiani et al.}

\graphicspath{ {../Figures/} }

\begin{document}

\title{Insight into the Formation of $\beta$ Pic b through the Composition of Its Parent Protoplanetary Disk as Revealed by the $\beta$ Pic Moving Group Member HD 181327}

\correspondingauthor{Henrique Reggiani}
\email{henrique.reggiani@noirlab.edu}

\author[0000-0001-6533-6179]{Henrique Reggiani}
\affiliation{Gemini Observatory/NSF's NOIRLab, Casilla 603, La Serena, Chile}
\affiliation{The Observatories of the Carnegie Institution
for Science, 813 Santa Barbara St, Pasadena, CA 91101, USA}

\author[0000-0001-9261-8366]{Jhon Yana Galarza}
\altaffiliation{Carnegie Fellow}
\affiliation{The Observatories of the Carnegie Institution
for Science, 813 Santa Barbara St, Pasadena, CA 91101, USA}

\author[0000-0001-5761-6779]{Kevin C.\ Schlaufman}
\affiliation{William H.\ Miller III Department of Physics and Astronomy, Johns Hopkins
University, 3400 N Charles St, Baltimore, MD 21218, USA}
\affiliation{Tuve Fellow, Carnegie Institution for Science Earth \& Planets Laboratory, 5241 Broad Branch Road NW, Washington, DC 20015, USA}

\author[0000-0001-6050-7645]{David K.\ Sing}
\affiliation{William H.\ Miller III Department of Physics and Astronomy, Johns Hopkins
University, 3400 N Charles St, Baltimore, MD 21218, USA}
\affiliation{Department of Earth and Planetary Sciences, Johns Hopkins
University, 3400 N Charles St, Baltimore, MD 21218, USA}

\author[0000-0002-7718-7884]{Brian F.\ Healy} 
\affiliation{School of Physics and Astronomy, University of Minnesota, 116 Church St SE, Minneapolis, MN 55455, USA}

\author{Andrew McWilliam}
\affiliation{The Observatories of the Carnegie Institution
for Science, 813 Santa Barbara St, Pasadena, CA 91101, USA}

\author[0000-0003-3667-8633]{Joshua D.\ Lothringer}
\affiliation{Department of Physics, Utah Valley University, College of
Science - MS 179, 800 W University Pkwy, Orem, UT 84058, USA}

\author[0000-0003-3818-408X]{Laurent Pueyo}
\affiliation{Space Telescope Science Institute, 3700 San Martin Drive, Baltimore, MD
21218, USA}

\begin{abstract}
\noindent
It has been suggested that $\beta$ Pic b has a supersolar metallicity and subsolar C/O ratio.  Assuming solar carbon and oxygen abundances for the star $\beta$ Pic and therefore the planet's parent protoplanetary disk, $\beta$ Pic b's C/O ratio suggests that it formed via core accretion between its parent protoplanteary disk's H$_{2}$O and CO$_{2}$ ice lines.  $\beta$ Pic b's high metallicity is difficult to reconcile with its mass $M_{\text{p}}~=~11.7~M_{\text{Jup}}$ though. Massive stars can present peculiar photospheric abundances that are unlikely to record the abundances of their former protoplanetary disks. This issue can be overcome for early-type stars in moving groups by infering the elemental abundances of the FGK stars in the same moving group that formed in the same molecular cloud and presumably share the same composition.  We infer the photospheric abundances of the F dwarf HD 181327, a $\beta$ Pic moving group member that is the best available proxy for the composition of $\beta$ Pic b's parent protoplanetary disk. In parallel, we infer updated atmospheric abundances for $\beta$ Pic b. As expected for a planet of its mass formed via core-accretion beyond its parent protoplanetary disk's H$_{2}$O ice line, we find that $\beta$ Pic b's atmosphere is consistent with stellar metallicity and confirm that is has superstellar carbon and oxygen abundances with a substellar C/O ratio. We propose that the elemental abundances of FGK dwarfs in moving groups can be used as proxies for the otherwise difficult-to-infer elemental abundances of early-type and late-type members of the same moving groups.

\end{abstract}


\section{Introduction}\label{intro}

It has been proposed that a giant planet's formation location in its parent protoplanetary disk can be discerned by studying the abundances
of the elements in the planet's atmosphere.  \citet{oberg2011} suggested that stellar C/O abundance ratios of giant planets 
can be used to infer which ice line (H$_2$O, CO, CO$_2$) 
its formation was interior to. Sub-stellar or stellar
C/O abundance ratios combined with superstellar carbon abundance
were argued to result from the accretion of large amounts of icy
planetesimals after envelope accretion. Superstellar C/O abundance
ratios and carbon abundances could be attributed either to formation
close to the CO$_{2}$ or CO ice lines or by the accretion of carbon-rich
grains in the narrow range inside the H$_{2}$O ice line but outside the
carbon-grain sublimation line.  Superstellar C/O abundance ratios and
sub-stellar carbon and oxygen abundances were put forward as a unique
signature of formation beyond the H$_{2}$O ice line. 

This straightforward scenario outlined for the case of static disks 
\citep{oberg2011} is significantly complicated in more realistic models that include
disk chemical and structural evolution, the radial migration of solids,
detailed models for planetesimal accretion, and/or post envelope-accretion
migration. It has also been argued that the abundances in giant
planet envelopes depend critically on the assumptions
made regarding the refractory composition of the inner disk. 
For a detailed discussion of the complications we refer the 
readers to Section $1$ of \citet{reggiani2022} and the 
papers cited therein 
\citep{harsono2015,piso2015,ali-dib2014,ali-dib2017,madhusudhan2014,mordasini2016,madhusudhan2017,booth2017,cridland2016,eistrup2018,cridland2019a,cridland2020,notsu2020,lothringer2021}. 

All of these analyses rely on two assumptions: (1) that the envelope of
a young giant planet stays well mixed during its formation even though
most metals are accreted before most gas and (2) that a mature giant
planet's atmosphere has a similar composition to the average composition
of its envelope at the end of the planet formation process.  Therefore, 
convection and mixing within the planetary atmosphere and planetary 
runaway during the gas accretion phase \citep[e.g.,][]{leconte2012,thiabaud2015} 
can also change the interpretation of 
these C/O ratios from the idealized view first proposed by \citet{oberg2011}. Despite all
of these complications, one robust prediction of giant planet formation
models in dynamically evolving disks is that the metal abundances of
giant planets with $M_{\text{p}} \lesssim 2~M_{\text{Jup}}$ are dominated
by the accretion of planetesimals after envelope accretion.  On the
other hand, the metal abundances of giant planets with $M_{\text{p}}
\gtrsim 2~M_{\text{Jup}}$ are dominated by envelope accretion itself
\citep{mordasini2014,mordasini2016,espinoza2017,cridland2019b}. 

The properties of the planet $\beta$ Pic b have been extensively studied 
in the literature. Previous studies indicate it has an estimated effective 
temperature of T$ =1590 \pm20$ K, 
a surface gravity log$(g/g_0)=4.0$, a metallicity of [Fe/H]$=0.5$ dex with a 
C/O$=0.43\pm0.05$ \citep{gravity2020}, a mass of $\textrm{M}=11.7\pm2.2 \ M_{Jup}$ 
\citep{2022ApJS..262...21F}, and an estimated radius of 
$\textrm{R}=1.46\pm0.01 \ R_{Jup}$ \citep{chilcote2017}. 
The analysis by the \cite{gravity2020} points to a slow formation via core-accretion, somewhere 
between the H$_2$O and CO$_2$ icelines. A scenario that can potentially explain the
subsolar C/O ratio if the planet was enriched in oxygen by icy
planetesimal accretion. For more details on the planet's formation, we refer the reader to 
\cite{gravity2020}.

However, the interpretation of giant planet formation is also dependent on its 
atmospheric carbon, oxygen, and C/O abundances, and as outlined above it is far from simple.  
Moreover, giant planet atmospheric
abundance ratios can only be meaningfully interpreted relative to the
mean compositions of their parent protoplanetary disk.  Directly inferring 
the abundances from protoplanetary disks is not straightforward 
even when the disk is still observable. However, in most cases the 
protoplanetary disk has already dissipated by the time we observe the planets. Therefore, 
the only way to reveal the mean compositions of those disks is
to use the photospheric abundances of their host stars.  During the era of
giant planet formation, the star growing at the center of a protoplanetary
disk has already accreted 99\% of the material that ever passed through
its disk.  As a result, the photospheric abundances of 
planet-host stars are an excellent proxy for mean protoplanetary disk
abundances.  The implication is that accurate and precise host star
elemental abundances for the same elements observed in giant planet
atmospheres are critically needed to achieve the full potential of giant
planet atmospheric abundance inferences as planet formation constraints, as 
was shown in \citet{reggiani2022}.

However, it is not always possible to directly infer all elemental abundances 
of a star. This is a notoriously difficult task for stars such as 
$\beta$ Pic because of its high effective temperature, surface gravity, and 
fast rotation 
\citep[$\textrm{T}_{eff}=8084\pm130$ K, log$g=4.22\pm0.13$, $v_{\textrm{micro}}=3.31$ km.s$^{-1}$, 
and vsin({i})$=113\pm1.13$ km.s$^{-1}$,][]{saffe2021}. Thus, it is not easy/possible to directly infer its 
entire chemical pattern. With exquisite HARPS spectra (R$\sim115000$, SNR$=1500$) 
\citet{saffe2021} was able to measure the abundances of \ion{C}{1}, \ion{Mg}{1}, \ion{Al}{1}, 
\ion{Si}{2}, \ion{Ca}{1}, \ion{Ca}{2}, \ion{Ti}{2}, \ion{Cr}{2}, \ion{Mn}{1}, \ion{Fe}{1}, 
\ion{Fe}{2}, \ion{Sr}{2}, \ion{Y}{2}, \ion{Zr}{2}, and \ion{Ba}{2}. Its oxygen 
abundance is noticeably missing, an essential element 
to the interpretation of planetary formation/migration 
scenarios, as described above. For $\beta$ Pic, to the best of  
our knowledge, oxygen was not yet measured \citep{gravity2020}.

There are extra complications to analyze $\beta$ Pic. 
It is a $\delta$ Scuti star \citep[e.g.,][]{mekarnia2017}, very close to the ZAMS \citep{crifo1997}, 
surrounded by a debris disk composed of dust and gas known to be continuously replenished by evaporating exocomets and colliding planetesimals 
\citep{ferlet1987,lecavelier1996,beust2000,wilson2017}. $\beta$ Pic has very low-amplitude periodic variations in brightness, radial
velocity, and line profiles \citep{koen2003a,koen2003b,galland2006}. 

Like many other A-type stars, $\beta$ Pic has a peculiar abundance pattern.
Typical abundance peculiarities of A-type stars can be as extreme as the 
large underabundances ($\sim1-2$ dex) of iron-peak elements, with 
near-solar abundances of C, N, O, and S seen in $\lambda$ Bo{\"o}tis stars 
\citep{kamp2001, andrievsky2002, heiter2002}. 
Up to 20\% of A- B-type stars have a wide range of chemical peculiarities, often associated 
with the presence of magnetic fields \citep[e.g.,][]{folsom2012}. 
\citet{folsom2012} also discusses the possibility that recently formed 
Hot Jupiters block the accretion of heavier material onto the star. 

Similar to other stars of its spectral type, 
$\beta$ Pic is also known to have a peculiar abundance pattern: 
sub-solar (in [X/H]) abundances of C, \ion{Mg}{1}, Al, Si, Sc, Ti, Cr, Mn, Fe, 
and Sr, while showing solar abundance of \ion{Mg}{2}, Y, and Ba, and super-solar 
abundance of Ca \citep{saffe2021}. $\beta$ Pic, 
formed $\sim30$ Myr ago in the thin disk of the Galaxy, 
does not follow the expected abundance pattern \citep[e.g.,][]{buder2021}. 
While the metallicity of $\beta$ Pic is [Fe/H]$=-0.28$ \citep{saffe2021}, 
the metallicity distribution function of the solar neighborhood peaks at 
[Fe/H]=0.0 \citep[e.g.,][]{kobayashi2020}. While it is not impossible for a young 
star to have the estimated metallicity of $\beta$ Pic, it is unlikely.

Therefore, as the abundance pattern of $\beta$ Pic, a star of a spectral type known to 
show chemical peculiarities (A-type), does not correspond to the expected abundance distribution 
of the solar neighborhood, the analysis of its photospheric abundances 
is unlikely to be a reliable tracer of the composition of the molecular cloud from which it 
formed. Therefore, it is also not a reliable tracer of the composition of the protoplanetary disk 
from which its planets were formed. On the other hand, stars are not formed 
isolated. They are formed in clusters, and stars that are co-natal 
share the same chemical composition. This assumption has been thoroughly tested in the literature, 
in studies testing the limits of chemical tagging \citep[e.g.][]{ness2018,andrews2019},
in analyzes of binary stars \citep[e.g.,][]{hawkins2020,nelson2021}, and stars in open 
clusters \citep[e.g.][]{ting2012,garcia-perez2016}.

In light of the above we propose that, for systems such 
as $\beta$ Pic, the stars that are part of their moving groups 
are a better probe of the interstellar material from which they, and 
their exoplanets, formed. In particular, the photospheric abundances of less massive, less evolved, 
stars in $\beta$ Pic's moving group are the best possible window into the 
protoplanetary disk's composition where $\beta$ Pic b formed. 

In this article we infer the potospheric and fundamental parameters 
as well as individual elemental abundances---including carbon and
oxygen---for a star that is part of $\beta$ Pic's moving group. We 
also perform a new retrieval of $\beta$ Pic b. We discuss how our 
results compare to the results previously published by \citet{gravity2020}. 
Our combined interpretation of the stellar abundances and planetary 
abundances corroborate the \citet{gravity2020} conclusions, although our results 
of star and planet are a better fit, especially for the planetary metallicity. 
We describe in Section \ref{data} the
high-resolution optical spectrum we collected for our programme stars.  We then
infer stellar parameters  in Section \ref{stellar_param}.  We derive the
individual elemental abundance in Section
\ref{elem_abund}. We present our retrieval of  $\beta$ Pic b in Section \ref{retrieval}. We review our results
and their implications in Section \ref{discussion}.  We conclude 
in Section \ref{conclusion}.

\section{Data}\label{data}

In order to analyze stellar data that would ultimately allow us to discuss $\beta$ Pic b's formation, we wanted to study stars that are part of $\beta$ Pic's moving group. In an effort to study the dynamical age of $\beta$ Pic's moving group, \citet{mireg2020} presented Gaia-based updated members of the moving group (see their Table 3). Their methodology is currently the gold standard for membership probability, i.e., they use the complete set of kinematic information available, provided by Gaia \citep{gaia2016,gaia2018}, to traceback the Galactic Orbit and associate individual stars with $\beta$ Pic's moving group. All the stars they identified as members of the moving group have a traceback position within 7 pc, with an uncertainty smaller than 2 pc at the time of minimum size. For reference, \cite{mireg2020} shows that the $\beta$ Pic association size (at time of birth) to be comparable to that of known starforming regions such as Ophiuchus \citep{canovas2019}, Taurus \citep{galli2019}, and Corona Australis \citep{galli2020}. For more details on the membership analysis we refer the reader to \cite{mireg2020}.

From their Table 3 (containing a total of 26 members of the $\beta$ Pic moving group), there are only two F type stars and two K type stars. The first K star is V{*} AO Men, an active eruptive variable. Its variability  makes it so that the analysis we aim to do is unreliable, as spectral lines can change based on variability yielding different stellar abundances that are dependent on the current level of stellar activity. Its temperature is also not ideal for the measurement of atomic carbon lines ($\mathrm{T_{eff}}=4431$ K based on an Infrared Flux Method estimative - \citealp{casagrande2021}). The second K star is CPD -72 2713. It is also active, and even cooler ($\mathrm{T_{eff}}=3992$ - based on the same analysis method); therefore, it is also not suitable for our methodology. Twenty-two out of the 26 members of the moving group are M stars. M stars have much more crowded optical spectra, leading to blended features. The spectra of M stars exhibit numerous molecular features, some of which are still being identified \citep[e.g.][]{crozet2023}, and a more complex atmospheric structure (for colder M dwarfs even clouds are an important factor). While we have many Benchmark analysis of FG stars, we are still benchmarking analysis methods of M dwarf \citep[e.g.,][]{souto2022,balmer2023}, which involves comparing their abundances to those of already benchmarked FGK dwarfs. In essence, all the M and K stars that are part of $\beta$ Pic's moving group are not amenable to accurate and precise analysis of their chemistry, and in particular of the carbon and oxygen abundances needed to interpret planetary formation and migration scenarios. Moreover, our analysis methodology \citep[e.g.,][]{reggiani2022} has been tested for FG and hot K stars. For these stars, we can obtain reliable stellar fundamental and photospheric parameters using our linelist and spectra+isochrones-based analysis.

Therefore, only two stars, HD 181327 and HD 191089, were suitable for our analysis methodology. We collected their spectra with the Magellan Inamori Kyocera Echelle (MIKE) spectrograph on the Magellan Clay Telescope at Las Campanas
Observatory \citep{bernstein2003,shectman2003}. We used the
0\farcs7~slit with standard blue and red grating azimuths,
yielding spectra between 335 nm and 950 nm with resolution $R \approx
40,\!000$ in the blue and $ R \approx 31,\!000$ in the red arms. 

We collected all calibration data (e.g., bias, quartz \& ``milky" flat field, and ThAr lamp frames) in the afternoon before each night of observations. We reduced the raw spectra and calibration frames using the \texttt{CarPy}\footnote{\url{http://code.obs.carnegiescience.edu/mike}} software package \citep{kelson2000,kelson2003,kelson2014}. Radial velocity correction and spectra normalization were performed using the Spectroscopy made Harder \citep[\texttt{smhr};][]{casey2014}\footnote{\url{https://github.com/andycasey/smhr/tree/py38-mpl313}} code.

\begin{figure}
\plotone{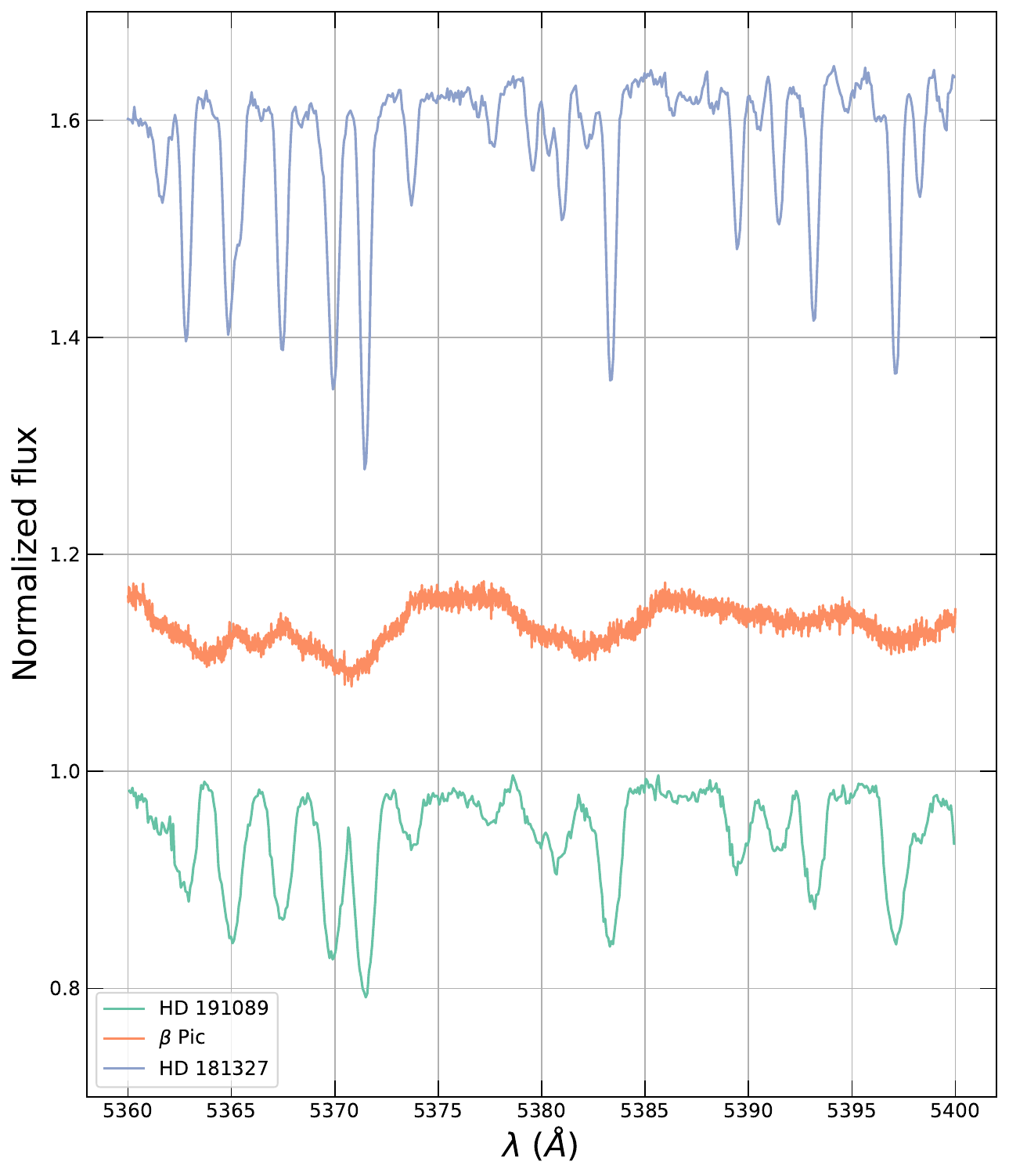}
\caption{We show the $4500-4540 \ \AA$ region of our spectra along with 
HARPS archival data of $\beta$ Pic from observing program ID 0104.C-0418(C). 
The high rotational velocities of HD 191089 \citep[$\approx 40$ km.s$^ {-1}$,][]{zuniga2021} 
and $\beta$ Pic are visible, and individual important atomic features are not possible to distinguish.
\label{specs}}
\end{figure}

Once our spectra were reduced and normalized we visually inspected them. After our inspection we dropped star HD 191089 from subsequent analysis. Because of the visibly high rotational velocity \citep[$\approx 40$ km.s$^ {-1}$,][]{zuniga2021}, some lines of interest (mainly carbon and oxygen transitions) are blended and cannot be reliably measured. In Figure \ref{specs} we show a region of the spectra of HD 181327, $beta$ Pic, and HD 191089. The high rotation of HD 191089, and the extreme velocity of $\beta$ Pic are easily observable. The $\beta$ Pic spectrum showed is an extremely high resolution (R$=115,000$), and high signal-to-noise (SNR$=215$) spectrum. Data for $\beta$ Pic was collected from the ESO archive, from observing program ID 0104.C-0418(C). After our data collection, reduction, and initial assessment, we will focus our efforts in studying HD 181327. This star is the only gold-standard confirmed $\beta$ Pic moving group member amenable to a never-before-published accurate and precise carbon and oxygen abundance inference, necessary for the interpretation of planetary formation scenarios.

\begin{deluxetable*}{lcccccccccc}
\tablecaption{Log of Magellan/MIKE Observations\label{log_obs}}
\tablewidth{0pt}
\tablehead{
\colhead{ID} & \colhead{R.A.} & \colhead{Decl.} & 
\colhead{UT Date} & \colhead{Start} & \colhead{Slit} & \colhead{Exposure} &
\colhead{$v_r$}  & \colhead{S/N} & \colhead{S/N} & \colhead{Spectral Type} \\
\colhead{Designation} & \colhead{(h:m:s)} & \colhead{(d:m:s)} &
\colhead{} & \colhead{} & \colhead{Width} & \colhead{Time (s)} & \colhead{(km s$^{-1}$)} &
\colhead{$4500 \ \rm{\AA}$} & \colhead{$6500 \ \rm{\AA}$} & \colhead{}
}
\startdata
HD 181327 & 19:22:58.94 & -54:32:16.98 & 2022 Apr 26 & 09:37:05 & 0\farcs7 & $70$ & $-21.8$ & $220$ & $335$ & F$7$\\
HD 191089 & 20:09:05.22 & -26:13:26.52 & 2022 Apr 28 & 09:16:23 & 0\farcs7 & $120$ & $-33.6$ & $300$ & $380$ & F$5^a$\\
\enddata
\tablecomments{a: Spectral type from \citet{mireg2020}}
\end{deluxetable*}

\section{Stellar Parameters}\label{stellar_param}

We derive photospheric and fundamental stellar parameters for our 
program star using the algorithm described in \citet{reggiani2022} that
makes use of both the classical spectroscopy-only approach\footnote{The
classical spectroscopy-only approach to photospheric stellar parameter
estimation involves simultaneously minimizing for individual line-based
iron abundance inferences the difference between \ion{Fe}{1} \&
\ion{Fe}{2}-based abundances as well as their dependencies on transition
excitation potential and measured reduced equivalent width.} and
isochrones to infer accurate, precise, and self-consistent photospheric 
(T$_{\text{eff}}$, log$g$, and [Fe/H)] and fundamental (mass, luminosity, and radius) 
stellar parameters.  

For our isochrone fitting 
we use high-quality multiwavelength photometry from the ultraviolet to the near-infrared: 
Tycho-2 BT and VT \citep{hog2000}, 
Gaia DR2 \citep{gaia2016,gaia2018,arenou2018,evans2018,hambly2018,riello2018, gaia2021,fabricius2021,lindegren2021a,lindegren2021b,torra2021} 
G, J, H, and Ks bands 
from the Two Micron All Sky Survey (2MASS) All-Sky Point Source Catalog 
\citep[PSC,][]{skrutskie2006}, and W1 and W2 bands from the 
Wide-field Infrared Survey Explorer (WISE) AllWISE mid-infrared
data \citep{wright2010,mainzer2011}.  We also include Gaia DR3 \citep{gaia2021} 
parallax-based distances \citep{bailer-jones2021} of our targets. 
We include the extinction $A_V$ inference based on three-dimensional
(3D) maps of extinction in the solar neighborhood from the Structuring
by Inversion the Local Interstellar Medium (Stilism) program
\citep{lallement2014,lallement2018,capitanio2017}.

For the spectroscopic-based inferences we use the equivalent widths (EWs) of \ion{Fe}{1} and \ion{Fe}{2} atomic absorption lines. The EWs were measured from our MIKE spectrum using Gaussian profiles with the \texttt{splot} task of IRAF\footnote{IRAF is distributed by the National Optical Astronomy Observatory, which is operated by the Association of the Universities for Research in Astronomy, Inc. (AURA) under cooperative agreement with the Na- tional Science Foundation.}.
The atomic absorption line data is an updated version from the lines 
described in \citet{galarza2019}. We assume \citet{asplund2021} photospheric solar abundances.

As described in detail in \citet{reggiani2022}, we use the \texttt{isochrones} 
package\footnote{\url{https://github.com/timothydmorton/isochrones}} \citep{morton2015} to fit the MESA Isochrones and Stellar Tracks \cite[MIST;][]{dotter2016,choi2016,paxton2011,paxton2013,paxton2015,paxton2018,paxton2019} library to our photospheric stellar parameters as well as our
input multiwavelength photometry, parallax, and extinction data using 
\texttt{MultiNest}\footnote{\url{https://ccpforge.cse.rl.ac.uk/gf/project/multinest/}} 
\citep{feroz2008,feroz2009,feroz2019} via \texttt{PyMultinest} \citep{buchner2014}.

We also analyzed the rotational period using TESS data and the technique of \citet{healy2020, healy2021, healy2023}. We processed PCA-detrended light curves from \texttt{eleanor} \citep{feinstein2019} by masking transits, removing outliers, flattening the light curves using 1-D polynomials, and binning the resulting points. We then generated a Lomb-Scargle periodogram and an autocorrelation function for each light curve, analyzing the peak positions and widths to determine periods and uncertainties.

Our adopted stellar parameters ($\rm{T_{eff}}$ and surface gravity from the isochrone analysis, 
and [Fe/H] and $\xi$ inferred from the atomic \ion{Fe}{1} and \ion{Fe}{2} lines)  are 
in Table \ref{stellar_param_table}. All of the uncertainties quoted in 
Table \ref{stellar_param_table} include random uncertainties only.  That is, they are 
uncertainties derived under the unlikely assumption that the MIST isochrone grid we use 
in our analyses perfectly reproduces all stellar properties.

\begin{deluxetable}{lcc}
\tablecaption{Adopted Stellar Parameters for HD 181327}\label{stellar_param_table}
\tablewidth{0pt}
\tablehead{
\colhead{Property} & \colhead{Value} & \colhead{Unit}}
\startdata
Tycho $B$  & $7.588\pm0.016$ & Vega mag \\
Tycho $V$  & $7.091\pm0.010$ & Vega mag \\
Gaia DR2 $G$ & $6.936\pm0.002$ & Vega mag \\
2MASS $J$ & $6.200\pm0.024$ & Vega mag \\ 
2MASS $H$ & $5.980\pm0.044$ & Vega mag \\ 
2MASS $K_{\text{s}}$ & $5.910\pm0.029$ & Vega mag \\ 
WISE W1 & $5.877\pm0.053$ & Vega mag \\ 
WISE W2 & $5.792\pm0.023$ & Vega mag \\ 
Gaia DR3 parallax & $20.931\pm0.029$ & mas \\
\hline
\multicolumn{3}{l}{\textbf{Isochrone-inferred parameters}} \\
Effective temperature $T_{\text{eff}}$ & $6498^{+56}_{-53}$ & K \\
Surface gravity $\log{g}$ & $4.33 \pm 0.01$ & cm s$^{-2}$ \\
Stellar mass $M_{\ast}$ & $1.36 \pm 0.02$ & $M_{\odot}$ \\
Stellar radius $R_{\ast}$ & $1.32\pm0.01$ & $R_{\odot}$ \\
Luminosity $L_{\ast}$ & $0.44 \pm 0.02$ & $L_{\odot}$ \\
Isochrone-based age $\tau_{\text{iso}}$ & $60\pm30$ & Myr \\
\hline
\multicolumn{3}{l}{\textbf{Spectroscopically inferred parameters}} \\
Metallicity $[\text{Fe/H}]$ &  $0.05\pm0.06$ & \\
Microturbulent velocity $\xi$ & $1.63 \pm 0.16$ & km s$^{-1}$ \\
\hline
\multicolumn{3}{l}{\textbf{Light-curve inferred parameters}} \\
Rotational period $P_{\text{rot}}$ & $1.542 \pm 0.048$ & day \\
\enddata
\end{deluxetable}

\section{Elemental Abundances}\label{elem_abund}

To infer the elemental abundances of the $\alpha$, light odd-$Z$, iron-peak, 
and neutron-capture elements we first measure the EWs of atomic absorption lines of \ion{Li}{1}, \ion{C}{1}, \ion{N}{1}, \ion{O}{1}, \ion{Na}{1}, \ion{Mg}{1}, \ion{Al}{1}, \ion{Si}{1}, \ion{S}{1}, \ion{K}{1}, \ion{Ca}{1}, \ion{Sc}{1}, \ion{Sc}{2}, \ion{Ti}{1}, \ion{Ti}{2}, \ion{V}{1}, \ion{Cr}{1}, \ion{Cr}{2}, \ion{Mn}{1}, \ion{Fe}{1}, \ion{Fe}{2}, \ion{Ni}{1}, \ion{Cu}{1}, \ion{Zn}{1}, \ion{Y}{2}, and \ion{Ba}{2} in our continuum-normalized spectrum by fitting Gaussian profiles with the \texttt{splot} task in \texttt{IRAF}.  We use the \texttt{deblend} task to disentangle absorption lines from adjacent spectral features whenever necessary. We measure an EW for every absorption line in our line list that could be recognized, taking into consideration the quality of the spectrum in the vicinity of a line and the availability of alternative transitions of the same species. We assume \citet{asplund2021} solar abundances and local thermodynamic equilibrium (LTE) and use the 1D plane-parallel solar-composition ATLAS9 \citep{Castelli2003} model atmospheres and the 2019 version of the LTE radiative transfer code \texttt{MOOG} to infer elemental abundances based on our EWs.

We report in Table \ref{elem_abundances} our abundance inferences 
in three common systems: $A(\text{X})$, $[\text{X/H}]$,
and $[\text{X/Fe}]$\footnote{$A(\text{X})=\log{N_{\text{X}}/N_{\text{H}}} + 12$\\
$[\text{X/H}] = A(\text{X}) - A(\text{X})_{\odot}$\\
$[\text{X/Fe}] = [\text{X/H}] - [\text{Fe/H}]$}.  We define 
the uncertainty in the abundance ratio $\sigma_{[\text{X/H}]}$ as the
square root of the sum of the square as the standard deviation of the
individual line-based abundance inferences $\sigma_{[\text{X/H}]}'$
divided by $\sqrt{n_{\text{X}}}$.  We define the uncertainty
$\sigma_{[\text{X/Fe}]}$ as the square root of the sum of squares of
$\sigma_{[\text{X/H}]}$ and $\sigma_{[\text{Fe/H}]}$.

\begin{deluxetable}{lrrcrcc}
\tablecaption{Elemental Abundances HD 181327}\label{elem_abundances}
\tabletypesize{\scriptsize}
\tablewidth{0pt}
\tablehead{
\colhead{Species} &
\colhead{$A(\text{X})$} &
\colhead{[X/H]} &
\colhead{$\sigma_{\text{[X/H]}}$} & 
\colhead{[X/Fe]} &
\colhead{$\sigma_{[\text{X/Fe}]}$} &
\colhead{$n$}}
\startdata
\multicolumn{6}{l}{LTE abundances} \\
\ion{Li}{1} & $3.49$ & $2.53$ & $\cdots$ & $2.48$ & $0.04$ & $1$ \\ 
\ion{C}{1} & $8.39$ & $-0.07$ & $0.06$ & $-0.12$ & $0.04$ & $5$ \\ 
\ion{N}{1} & $\le7.81$ & $\le-0.02$ & $\cdots$ & $\le-0.07$ & $\cdots$ & $1$ \\ 
\ion{O}{1} & $8.92$ & $0.23$ & $0.09$ & $0.18$ & $0.08$ & $3$ \\ 
\ion{Na}{1} & $6.20$ & $-0.02$ & $0.12$ & $-0.07$ & $0.12$ & $2$ \\ 
\ion{Mg}{1} & $7.65$ & $0.10$ & $0.12$ & $0.04$ & $0.07$ & $5$ \\ 
\ion{Al}{1} & $6.21$ & $-0.22$ & $0.34$ & $-0.28$ & $0.24$ & $3$ \\ 
\ion{Si}{1} & $7.77$ & $0.26$ & $0.32$ & $0.20$ & $0.10$ & $12$ \\ 
\ion{S}{1} & $7.20$ & $0.08$ & $0.26$ & $0.02$ & $0.19$ & $3$ \\ 
\ion{K}{1} & $5.64$ & $0.57$ & $0.00$ & $0.52$ & $0.07$ & $1$ \\ 
\ion{Ca}{1} & $6.54$ & $0.24$ & $0.19$ & $0.18$ & $0.07$ & $11$ \\ 
\ion{Sc}{1} & $3.53$ & $0.39$ & $0.38$ & $0.33$ & $0.38$ & $2$ \\ 
\ion{Sc}{2} & $3.57$ & $0.43$ & $0.61$ & $0.38$ & $0.22$ & $9$ \\ 
\ion{Ti}{1} & $5.23$ & $0.26$ & $0.38$ & $0.21$ & $0.12$ & $12$ \\ 
\ion{Ti}{2} & $5.49$ & $0.52$ & $0.88$ & $0.47$ & $0.32$ & $9$ \\ 
\ion{V}{1} & $4.54$ & $0.64$ & $0.05$ & $0.59$ & $0.07$ & $2$ \\ 
\ion{Cr}{1} & $5.74$ & $0.12$ & $0.24$ & $0.07$ & $0.09$ & $11$ \\ 
\ion{Cr}{2} & $5.90$ & $0.28$ & $0.21$ & $0.23$ & $0.10$ & $7$ \\ 
\ion{Mn}{1} & $5.26$ & $-0.16$ & $0.30$ & $-0.22$ & $0.13$ & $7$ \\ 
\ion{Fe}{1} & $7.51$ & $0.05$ & $0.09$ & $\cdots$ & $\cdots$ & $37$ \\ 
\ion{Fe}{2} & $7.63$ & $0.17$ & $0.15$ & $\cdots$ & $\cdots$ & $15$ \\ 
\ion{Ni}{1} & $6.34$ & $0.14$ & $0.34$ & $0.09$ & $0.09$ & $17$ \\ 
\ion{Cu}{1} & $4.35$ & $0.17$ & $0.29$ & $0.11$ & $0.29$ & $2$ \\ 
\ion{Zn}{1} & $4.51$ & $-0.05$ & $0.08$ & $-0.11$ & $0.10$ & $2$ \\ 
\ion{Y}{2} & $2.70$ & $0.49$ & $0.62$ & $0.43$ & $0.36$ & $4$ \\ 
\ion{Ba}{2} & $2.79$ & $0.52$ & $0.22$ & $0.47$ & $0.15$ & $4$ \\ 
\hline
\multicolumn{7}{l}{1D non-LTE abundances$^a$} \\
\ion{C}{1} & $8.37$ & $-0.09$ & $0.06$ &$-0.19$ & $\cdots$ & $1$ \\ 
\ion{O}{1} & $8.56$ & $-0.13$ & $0.09$ &$-0.23$ & $\cdots$ & $1$ \\ 
\ion{Al}{1} & $6.10$ & $-0.33$ & $0.36$ &$-0.43$ & $\cdots$ & $2$ \\ 
\ion{Si}{1} & $7.74$ & $0.23$ & $0.30$ &$0.13$ & $\cdots$ & $12$ \\ 
\ion{K}{1} & $5.14$ & $0.07$ & $0.00$ &$-0.3$ & $\cdots$ & $1$ \\ 
\ion{Ca}{1} & $6.04$ & $-0.26$ & $0.06$ &$-0.36$ & $\cdots$ & $2$ \\ 
\ion{Fe}{1} & $7.56$ & $0.10$ & $0.09$ &$\cdots$ & $\cdots$ & $37$ \\ 
\ion{Fe}{2} & $7.68$ & $0.22$ & $0.16$ &$\cdots$ & $\cdots$ & $15$ \\ 
\hline
\multicolumn{7}{l}{3D non-LTE abundances} \\
\ion{Li}{1} & $3.27$ & $2.31$ & $\cdots$ & $2.18$ & $0.04$ & $1$ \\ 
\ion{C}{1} & $8.38$ & $-0.08$ & $0.06$ &$-0.18$ & $\cdots$ & $5$ \\ 
\ion{O}{1} & $8.59$ & $-0.10$ & $0.06$ &$-0.20$ & $\cdots$ & $3$ \\ 
\hline
\multicolumn{7}{l}{Additional abundance ratios of interest$^b$} \\
\multicolumn{7}{l}{$[\text{Fe/H}]_{\text{1D non-LTE}} = +0.13 \pm 0.10$} \\
\multicolumn{7}{l}{$[\text{(C+O)/H}]_{\text{1D LTE}} = +0.16 \pm 0.11$} \\
\multicolumn{7}{l}{$[\text{(C+O)/H}]_{\text{1D non-LTE}} = -0.22 \pm 0.11$} \\
\multicolumn{7}{l}{$[\text{(C+O)/H}]_{\text{3D non-LTE}} = -0.18 \pm 0.08$} \\
\multicolumn{7}{l}{$[\text{C/O}]_{\text{1D LTE}} = -0.3 \pm 0.11$} \\
\multicolumn{7}{l}{$[\text{C/O}]_{\text{1D non-LTE}} = +0.04 \pm 0.11$} \\
\multicolumn{7}{l}{$[\text{C/O}]_{\text{3D non-LTE}} = +0.02 \pm 0.08$} \\
\multicolumn{7}{l}{C/O$_{\text{1D LTE}} = +0.30 \pm 0.11$}\\
\multicolumn{7}{l}{C/O$_{\text{1D non-LTE}} = +0.65 \pm 0.11$}\\
\multicolumn{7}{l}{C/O$_{\text{3D non-LTE}} = +0.62 \pm 0.08$}
\enddata
\tablenotetext{a}{[X/Fe] ratios calculated using the 1D non-LTE corrected \ion{Fe}{1} a
bundances ([\ion{Fe}{1}/H]$=+0.10$).}
\tablenotetext{b}{C/O ratios are defined as C/O$=10^{(A(C) - A(O))}$.}
\end{deluxetable}

When possible, we update our elemental abundances for departures from 
LTE (i.e., non-LTE corrections) by linearly interpolating published grids
of non-LTE corrections using \texttt{scipy} \citep{virtanen2020}.
We use the publicly available correction tables from \citet{amarsi2020} for 
carbon, oxygen, aluminium, and calcium. Silicon corrections are from 
\citet{amarsi2017}, potassium corrections are from \citet{reggiani2019}, 
and iron corrections are from \citet{amarsi2016}.  
There is one caveat for our carbon 1D non-LTE 
corrected abundance. The \citet{amarsi2020} non-LTE grid 
only includes one carbon transition: the $6588 \ \AA$ transition. This line 
is visible in our spectrum, but blended with a telluric feature which 
we have not been able to properly remove. Therefore, we used 
the mean 1D LTE abundance of all carbon lines (Table \ref{elem_abundances}) 
to estimate our correction. For oxygen, the grid has a mean correction based on 
all three lines at the $777$ nm region. We, therefore, use our mean 1D LTE 
oxygen abundance (Table \ref{elem_abundances}) to estimate the correction. 
Our reported 3D non-LTE abundances for carbon and oxygen are a nearest neighbor 
extrapolation based on the \citet{amarsi2019} grid. Our extrapolations 
are to values extremely close to the edge of the correction grids, and therefore reliable results. 
All our non-LTE estimated abundances are presented in Table \ref{elem_abundances}. 

We estimated possible systematics in our non-LTE corrections by interpolating 
oxygen abundance corrections from \citet{bergemann2021}, through the 
online tool \textit{Spectrum Tools}\footnote{\href{https://nlte.mpia.de/}{https://nlte.mpia.de/}}. 
Our interpolation resulted in a mean 1D non-LTE correction of $\Delta\text{A(O)}=-0.18$, 
considerably different than the 1D non-LTE correction we inferred ($\Delta\text{A(O)}=-0.36$) from \citet{amarsi2020}. 

We confirmed our own estimates by interpolating the same corrections using the online tool 
INSPECT\footnote{\href{http://www.inspect-stars.com/}{http://www.inspect-stars.com/}}, 
which is based on the same oxygen model-atom used in \cite{amarsi2020}. For its interpolation 
process \textit{Spectrum Tools} does not request line abundances, therefore the interpolated 
correction is not specific for a line/abundance measurement. It does, however, 
show the curve of growth from which the result was interpolated. From visual inspection of the curves 
of growth, the abundance corrections appear higher than what we reported here ($\Delta\text{A(O)}=-0.18$).
Nevertheless, because of the uncertainty generated by not being able to directly interpolate 
our abundances, we chose to adopt the \citet{amarsi2020}, \citet{amarsi2019} 1D and 3D corrections, respectively. 

Even so, we caution the reader of a possible systematic in our oxygen corrections of up to $\approx0.18$ dex. 
It is outside the scope of this study to understand the differences between the models that might 
be responsible for this difference, but we point that if we were to adopt the 1D non-LTE correction from 
\textit{Spectrum Tools}, the C/O ratio would be C/O$_{\text{1D non-LTE}}=+0.43$. This different C/O ratio would not, however, 
change the interpretations of our results (Section \ref{betapicb_form}). 

\begin{deluxetable*}{cccccc}
\tablecaption{Atomic data, Equivalent Widths and line Abundances. Full version online.\label{measured_ews}}
\tablewidth{0pt}
\tablehead{
\colhead{Wavelength} & \colhead{Species} &
\colhead{Excitation Potential} & \colhead{log($gf$)} &
\colhead{EW} & \colhead{$\log_\epsilon(\rm{X})$} \\ 
 & \colhead{(\AA)} &  & \colhead{(eV)} & & (m\AA)  }
\startdata
$6154.225$ & \ion{Na}{1} & $2.102$ & $-1.547$ & $25.3$ & $6.318$\\ 
$6160.747$ & \ion{Na}{1} & $2.104$ & $-1.246$ & $28.3$ & $6.081$\\ 
$4571.095$ & \ion{Mg}{1} & $0.000$ & $-5.623$ & $87.5$ & $7.781$\\ 
$4730.040$ & \ion{Mg}{1} & $4.340$ & $-2.389$ & $44.0$ & $7.610$
\enddata
\tablecomments{This table is published in its entirety in the machine-readable format.  A portion is shown here for guidance regarding its form and content.} 
\end{deluxetable*}

\section{$\beta$ Pic b Retrieval}
\label{retrieval}

\subsection{Retrievals for $\beta$ Pic b}
We performed spectral retrievals on $\beta$ Pic b, using the GRAVITY and GPI data 
\citep{gravity2020, chilcote2017}. Compared to previous work, we re-calibrated the GPI data following the procedures in \citet{DeRosa2020SPIE11447E..5AD} which improves the photometric precision and reduces systematic errors.
We use \texttt{ATMO}, a 1D-2D radiative-convective equilibrium model for planetary atmospheres to generate models of beta Pic b. More comprehensive descriptions of the model can be found in \citet{2014A&A...564A..59A, 2015ApJ...804L..17T, 2016ApJ...817L..19T, 2017ApJ...841...30T, 2016A&A...594A..69D},  \citet{2018MNRAS.474.5158G}, and \citet{2018MNRAS.474.5158G}. 

\begin{deluxetable}{ll}
\tablecaption{\texttt{ATMO} Retrieval parameters.\label{tab_retrieval}}
\tablewidth{0pt}
\tablehead{\colhead{Parameter} & \colhead{Posterior}}
\startdata
\hline
\multicolumn{2}{l}{Fit parameters} \\
$[Z/Z$$_{\odot}]$                & 0.021$_{-0.117} ^{+0.243}$\\
$R$$_{pl}$(R$_{\rm Jup}$,1 mbar)     & 1.2941$_{-0.011}^{+0.015}$\\  
cloud opacity ln($\delta$)       & 3.26$_{-0.48}^{+0.32}$\\  
log cloud top (bar)              & -0.982$_{-0.162}^{+0.115}$\\  
$\kappa_{IR}$ (cm$^2$/g)         & -1.29$\pm$-0.17\\ 
$\gamma_1$                       & -3.57$_{-1.52}^{+1.37}$\\  
Internal Temp (K)                & 1840$_{-18}^{+12}$\\
$[\text{C/H}]$           & 0.255$_{-0.150}^{+0.179}$\\
$[\text{O/H}]$           & 0.467$_{-0.158}^{+0.159}$ \\
$\text{A(C)}$  &  $8.755_{-0.150}^{+0.179}$ \\
$\text{A(O)}$  &  $9.227_{-0.158}^{+0.159}$ \\
\hline
\multicolumn{2}{l}{\citet{asplund2021} Scaled Abundances} \\
$[\text{Fe/H}]$ & $0.081^{+0.246}_{-0.124}$\\
$[\text{C/H}]$           & $0.295_{-0.155}^{+0.183}$\\
$[\text{O/H}]$           & $0.537_{-0.163}^{+0.164}$ \\
\hline
\multicolumn{2}{l}{Inferred parameters} \\
C/O                              & 0.35$_{-0.03}^{+0.02}$\\
Mass ($M_{\rm Jup}$)             & 7.2$_{-2.5}^{+3.6}$
\enddata
\end{deluxetable}

\begin{figure*}
\plotone{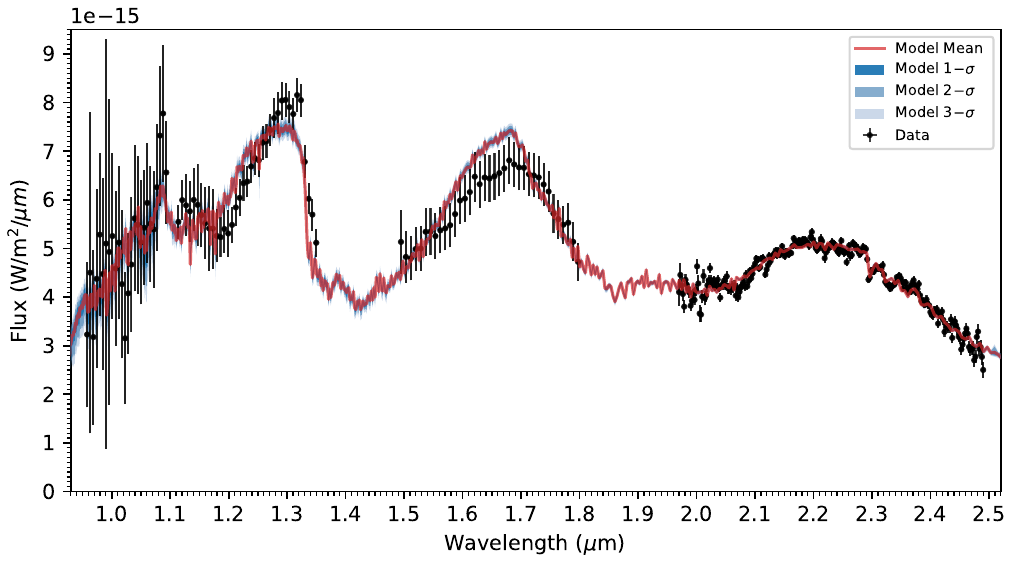}
\caption{$\beta$ Pic b GPI and Gravity spectrum (black datapoints) along with the posterior distribution of the spectral retrieval model. \label{atmo_spec}}
\end{figure*}

For the model assuming chemical equilibrium, the elemental abundances for each model were freely fit and calculated in equilibrium on the fly.  
Two elements were selected to vary independently, as they are major species which are also likely to be sensitive to spectral features in the data, while the rest were varied by a metallicity parameter ([Z/Z$_{\odot}$]).  By varying the carbon and  oxygen,elemental abundances separately ([C/C\textsubscript{$\odot$}] \& [O/O\textsubscript{$\odot$}]) we allow for non-solar compositions but with chemical equilibrium imposed such that each model fit has a chemically-plausible mix of molecules given the retrieved temperatures, pressures and underlying elemental abundances. 
In addition, we alleviate an important modeling assumption which can affect the retrieved C/O value (see \citealt{2019MNRAS.486.1123D}). 
The solar abundances of C, N, O, P, S, K, Fe were defined from \citet{Caffau2011SoPh..268..255C} while we used \citet{asplund2009} for the other species. 
As our stellar abundances are scaled to the \citet{asplund2021} photospheric solar abundances we also present, in Table \ref{tab_retrieval}, the planetary metallicity, as traced by iron, carbon, and oxygen abundances re-scaled to the \citet{asplund2021} solar photospheric abundances. While the solar abundance uncertainties were not included in our original retrieval, we include the uncertainties of the \citet{asplund2021} solar abundances in Table \ref{tab_retrieval}. The updated uncertainties are the square root of the sum of squares of the planetary uncertainties and the \citet{asplund2021} uncertainties. 
For the spectral synthesis, we included the spectroscopically active molecules of H$_{2}$, He, H$_{2}$O, CO$_{2}$, CO, CH$_{4}$, NH$_{3}$, H$_2$S, HCN, C$_2$H$_2$, Na, K, TiO, VO, FeH, Fe, and H-.  
To parameterize the T-P profile, we use the analytic radiative equilibrium model as described \citet{Guillot2010A&A...520A..27G}. We vary three parameters, corresponding to one optical channel, one infrared channel and the internal temperature. The suface gravity was also allowed to vary.

\begin{figure*}
\includegraphics[scale=0.32]{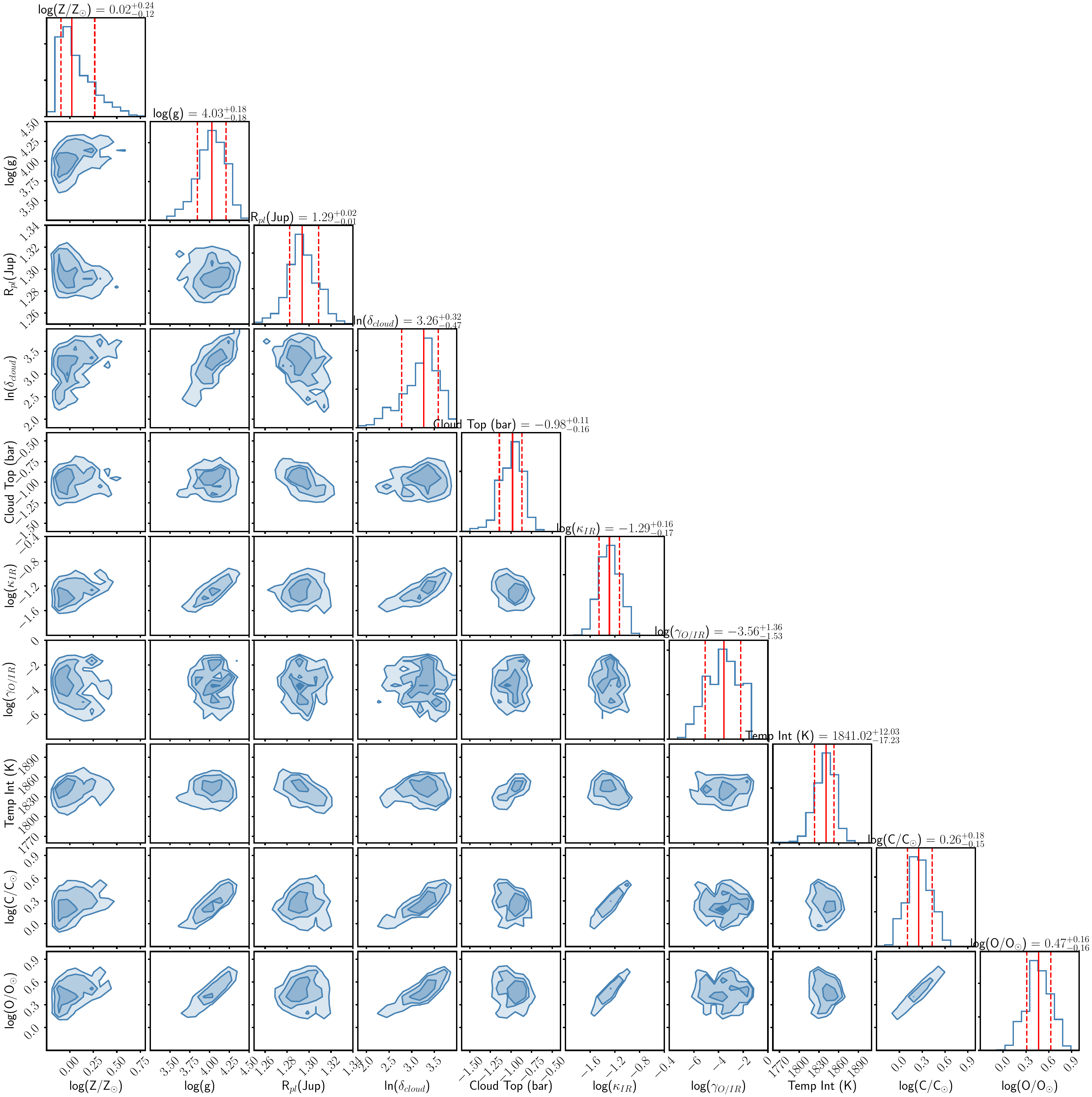}
\caption{Posterior distribution for the \texttt{ATMO} atmospheric retrieval model fit to the $\beta$ Pic b spectrum. \label{atmo_corner}}
\end{figure*}

We also included a grey cloud parameterized by a uniform opacity and a cloud top pressure level. 
Rainout of condensate species was also included.  In total, 10 free parameters were used to fit the 337 datapoints. The results are shown in Figs. \ref{atmo_spec} and \ref{atmo_corner} with the results also given in Table \ref{tab_retrieval} including the best-fitting model parameters and 1 $\sigma$ uncertainties.  We retrieve a mass of 7.2$_{-2.5}^{+3.6}$ $M_{Jup}$ which consistent with the latest estimates of 9 to 11 $M_{Jup}$ using radial velocity, astrometry, and direct imaging data \citet{2022ApJS..262...21F, 2019NatAs...3.1135L}.

\section{Discussion}
\label{discussion}

\subsection{HD 181327 as a Tracer of the Composition of $\beta$ Pic b's Proto-Planetary Disk}
\label{}

The inferred metallicity of $\beta$ Pic is subsolar \citep{saffe2021}. Although technically not impossible, it is very unlikely for a solar neighborhood, younger than 100 Myrs star, to be formed in a low metallicity environment \citep[e.g.][]{kobayashi2020}. On the other hand, the chemical abundances we inferred for HD 181327 are within the expected range of abundances for a solar neighborhood, solar/mildly super-solar metallicity star \citep[both observationally and theoretically,][]{kobayashi2020}, unlike $\beta$ Pic itself. This difference is a result of $\beta$ Pic's chemical peculiarities, (see Section \ref{intro} and the brief discussion on A stars). This provides initial, albeit indirect, evidence that HD 181327 might be a better representation of the composition of the molecular cloud from which $\beta$ Pic, and $\beta$ Pic b formed.

Stars are formed in clusters, and moving groups are remnants of the clusters where stars were formed. There is now ample evidence indicating that stars formed together are chemically homogeneous. Previous studies of co-moving pairs of stars show that their abundances are homogeneous at the $0.05$ dex level \citep[e.g.,][]{hawkins2020,nelson2021}.

Studying open clusters homogeneity, \citet{pooveli2020} recently found that their chemistry is homogeneous at the $0.03$ dex level. They used APOGEE to study a set of ten open clusters. Each open cluster has at least 13 members (ranging from 13 to 381 members), with an average of 29 members per cluster, excluding the one cluster with 381 studied members (the only cluster with more than 60 members in their sample). This extremely good homogeneity has previously been reported in other open clusters as well: IC 4756 has a star-to-star variation smaller than [X/H]$<0.03$ dex \citep[e.g.,]{ting2012}; The Hyades are homogeneous within $0.02$ dex \citep{desilva2006,desilva2011,liu2016a}; M67, NGC 6819, and NGC 2420 also show homogeneity within $0.03$ dex \citep{bovy2016}; \citet{bertran2016} found homogeneity better than $0.01$ dex for [O/Fe] in several clusters. In conclusion, it is safe to assume that open clusters are homogeneous. Consequently, we can assume that stars in the same moving group are chemically homogeneous at least to the $0.03$ dex level. 

\citet{mireg2020} found 26 members of $\beta$ Pic's moving group, a similar number of members as those of the \citet{pooveli2020} open clusters study. With a number of members similar to the previously studied clusters, we assume the homogeneity of the $\beta$ Pic moving group to be similar (dispersion $<0.03$ dex). Therefore, if one can measure the chemical abundances of a reliable tracer of the moving group, its abundances should be within the order of $\sim 0.03$ dex of the remaining moving group members. Thus, the abundances inferred for HD 181327 should be a good reflection of the abundances of the molecular cloud where the $\beta$ Pic moving group formed.

\begin{figure}
\plotone{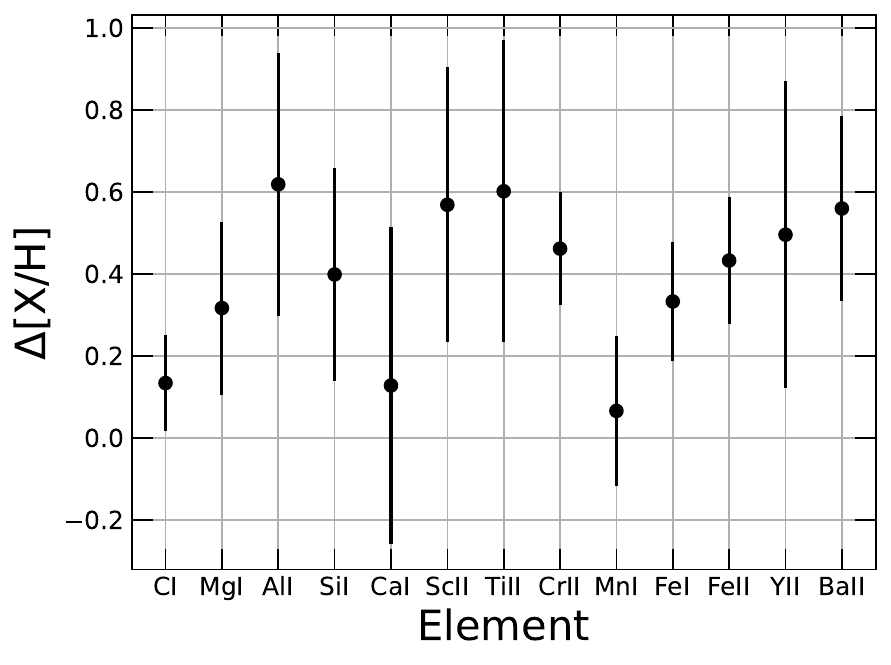}
\caption{We show the abundance differences between HD 181327 and $\beta$ Pic. The abundances from $\beta$ Pic are from \citet{saffe2021}.\label{abnd_comp}}
\end{figure}

That is not the case for $\beta$ Pic itself. It is an A star with an unusual chemical composition: sub-solar (in [X/H]) abundances of C, Mg, Al, Si, Sc, Ti, Cr, Mn, Fe, and Sr, while showing solar abundance of \ion{Mg}{2}, Y, and Ba, and super-solar abundance of Ca \citep{saffe2021}. $\beta$ Pic is also classified as a $\delta$ Scuti star, with over 30 $\delta$ Scuti pulsation frequencies identified \citep{mekarnia2017}. The pulsation alone makes it challenging for an abundance analysis, as spectroscopic line-profiles vary with pulsation \citep{aerts2008}. Moreover, its high rotational velocity ($>100 \ \mathrm{km.s^{-1}}$, see Fig. \ref{specs}) makes it challenging for an accurate and precise determination of several species, such as oxygen (never directly inferred from $\beta$ Pic). In fact, because of the difficulties, and intrinsic characteristics of $\beta$ Pic, a comparison between its composition and the composition we inferred for HD 181327 shows differences far exceeding the $0.03$ dex homogeneity observed in stars formed within the same molecular clouds. We show these differences in Figure \ref{abnd_comp}. Because the inferred chemistry of $\beta$ Pic cannot be reliably used as a proxy for its molecular cloud (due to its pulsation, rotational velocity, and chemical peculiarities), we turned ourselves to another proxy: Hd 181327.

Based on the information presented above, we argue that, for the specific case of $\beta$ Pic b, the best way to interpret its planetary retrieved chemistry is to analyze the chemical makeup of star HD 181327, the only slow rotator, low activity, F star in $\beta$ Pic's moving group amenable to the analysis we perform in this study, and for which both carbon and oxygen (among other elements) can be accurately inferred. When it comes to the remaining members of $\beta$ Pic's moving group, to the best of our knowledge, there are no non-LTE, or 3D, carbon and oxygen abundance corrections available for M stars, and any analysis are therefore not as accurate as that of F stars. Thus, not only the analysis of HD 181327 is a better representation of $\beta$ Pic b's protoplanetary disk than $\beta$ Pic, but also the only analysis that allows us to accurately and precisely infer the abundances of key species such as carbon and oxygen in a well understood stellar context. 

\begin{figure*}
\plottwo{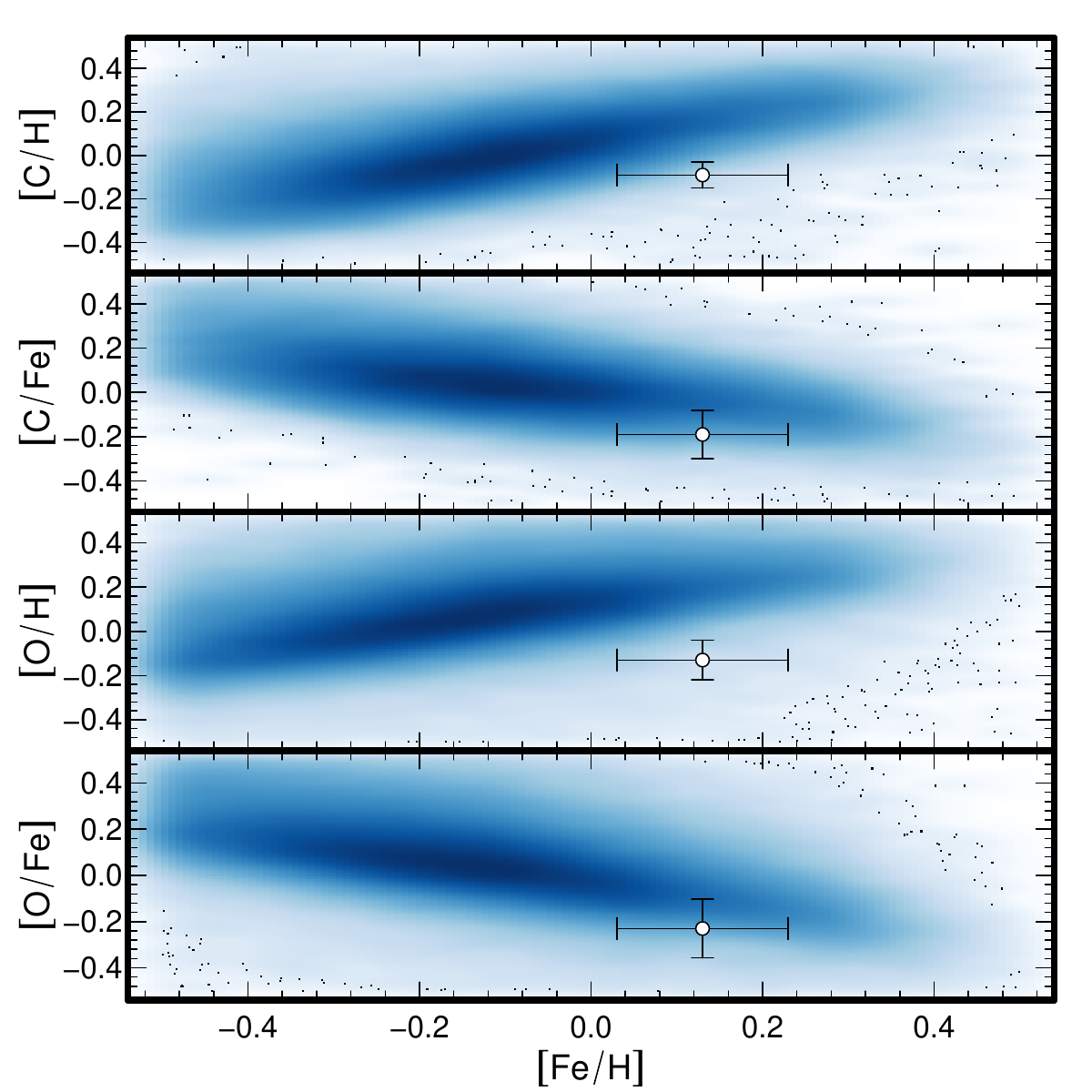}{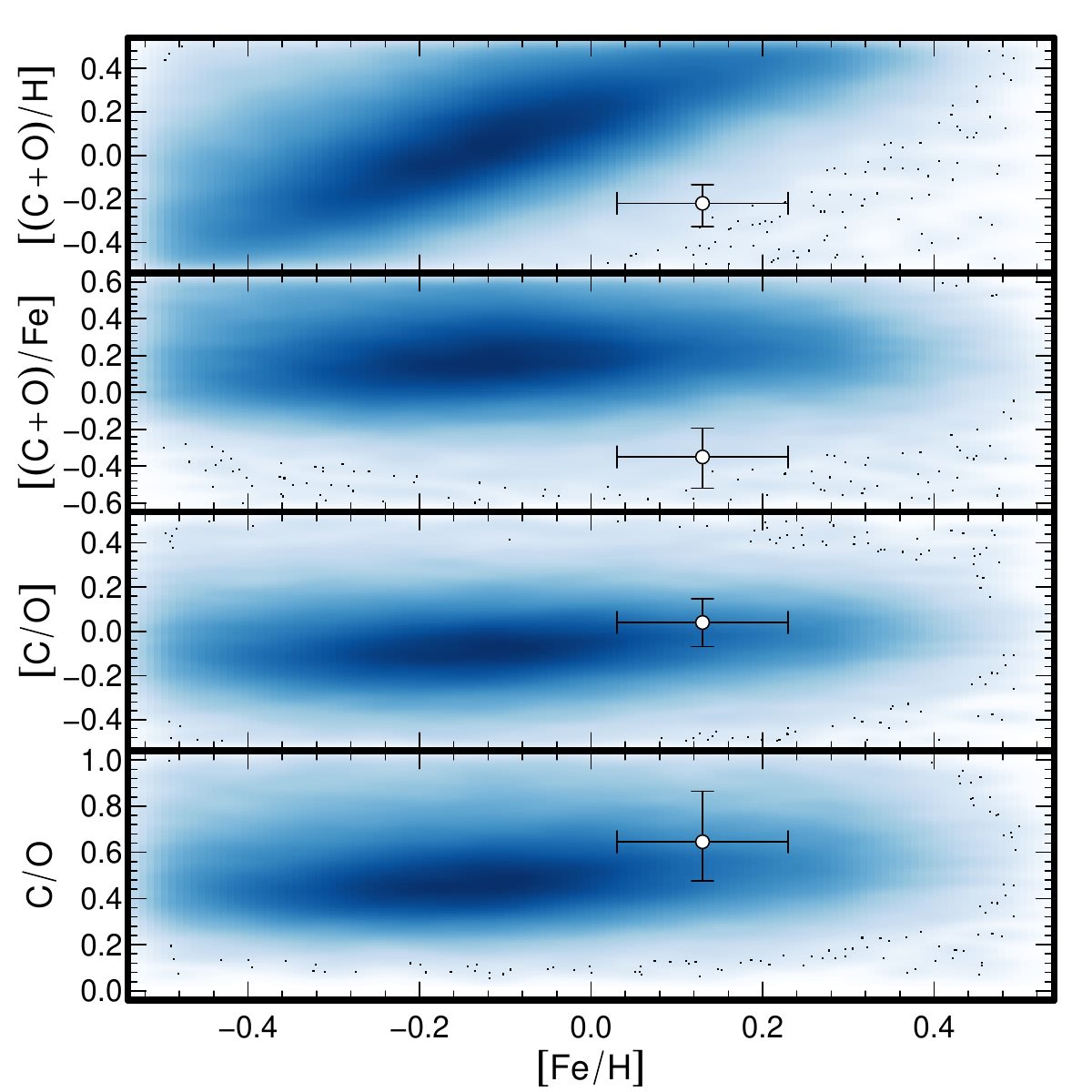}
\caption{We present, as blue points, the distribution of carbon and oxygen abundances 
(in different scales) as a function of metallicity 
from the GALAH DR3 survey. In white we show the abundance of HD 181327, as a proxy for 
$\beta$ Pic's molecular cloud. One can see that its abundance is outside of the most populated region 
in all abundance spaces. This shows the importance of individually determining the stellar abundances for 
the interpretation of planetary retrievals.\label{galah_plot}}
\end{figure*}

\citet{gravity2020} made an extensive discussion of $\beta$ Pic b formation 
scenario based on the 
assumption that $\beta$ Pic's C/O ratio is 
solar (C/O$=0.59$, from \citealp{asplund2021}), and according to 
them: A subsolar C/O ratio (i.e., $\approx$ 0.4) would
invalidate most of the discussion regarding the formation 
of $\beta$ Pic b. In Figure \ref{galah_plot} we 
show the distribution of carbon and oxygen ratios in the solar neighborhood, using data 
from the third data release of the Galactic Archaeology with HERMES (GALAH) survey 
\citep{buder2021}, and in white the abundances of HD 181327. It is clear from Figure \ref{galah_plot} 
that the abundances of HD 181327 are not located in the bulk of the solar neighborhood distribution. 
From GALAH, the peak of the distribution of C/O ratios at solar metallicity, in the solar 
neighborhood, is $\sim0.1$ dex below solar (C/O$=0.49^{+0.15}_{-0.12}$). Thus, 
for a solar neighborhood, solar metallicity star, it is more likely to find subsolar than solar 
abundance for the C/O ratio, and this fact must be taken into account when analyzing planetary 
abundance information.

As one needs to interpret planetary formation and migration processes taking into 
account the composition of the disk from which the planet formed \citep[e.g.,][]{reggiani2022}, 
we propose that the interpretation of planetary abundances should be done in light of:

\begin{enumerate}
\item Preferably through the photospheric abundances of the host star.
\item The photospheric abundance of a star formed within the same molecular cloud 
as the host star/planet. 
\item The mean abundances observed in the Galactic substructure where we observe 
the planet (solar neighborhood).
\item If all other options are not possible, one should assume solar abundances.
\end{enumerate}

For the photosphere of HD 181327, and therefore the protoplanetary disk from 
which $\beta$ Pic b formed, we found (3D non-LTE corrected abundances)  
[C/H]$=-0.08\pm0.06$, [O/H]$=-0.10\pm0.06$, [(C+O)/H]$=-0.18\pm0.08$, C/O]$=+0.02\pm0.08$, and C/O$=+0.62\pm0.08$. We will discuss $\beta$ Pic b's formation using the 3D non-LTE abundances calculations as the basis of our discussions. We reinforce that our carbon 1D LTE, 1D non-LTE, and 3D non-LTE abundances are all compatible, and the differences between them are smaller than the uncertainties.

\subsection{The Formation of $\beta$ Pic b}
\label{betapicb_form}

Our retrieval of $\beta$ Pic b yields different results than those reported by the \citet{gravity2020}. We attribute the differences mostly to our updated methodology in the GPI data reduction. Compared to the \citet{gravity2020} retrieval, our metallicities are in much better agreement to the stellar results. The 1D non-LTE metallicity of the planetary disk, as traced by HD 181327, is [Fe/H]$_{\text{1D NLTE}}=+0.13\pm0.10$, and the planetary metallicity is [Fe/H]$=0.5$, and [Fe/H]$=+0.08^{+0.25}_{-0.12}$ from \citet{gravity2020} and our own retrieval, respectively. Ours is a clear improvement as it is now fully reconciled to the metallicity traced by the stellar information recorded in HD 181327. Not only that, but the much lower metallicity we inferred is much more amenable to the core-accretion formation scenario of $\beta$ Pic $b$, than the [Fe/H]$=0.5$ dex previously inferred.

Our retrieved C/O abundance ratio is smaller than what was previously reported, 
although there is agreement within the quoted uncertainties (C/O$=+0.35^{+0.02}_{-0.03}$ 
from us and C/O$=0.43\pm0.05$ from \citet{gravity2020}). The \citet{gravity2020} team, however, 
did not report the individual carbon and oxygen abundances, as carbon in their retrieval was 
scaled to the planetary metallicity\footnote{\href{https://petitradtrans.readthedocs.io/en/latest/content/notebooks/poor_man.html}{See the petitRADTRANS documentation for details.}}, 
and oxygen was freely varied to recover the inferred C/O ratio. 
In our retrieval the abundances are estimated on the fly and every element retrieved is 
treated as a free parameter. Therefore, our carbon and oxygen abundances are not biased by 
the assumption of solar scaled [C/H]. 

\begin{figure*}
\plottwo{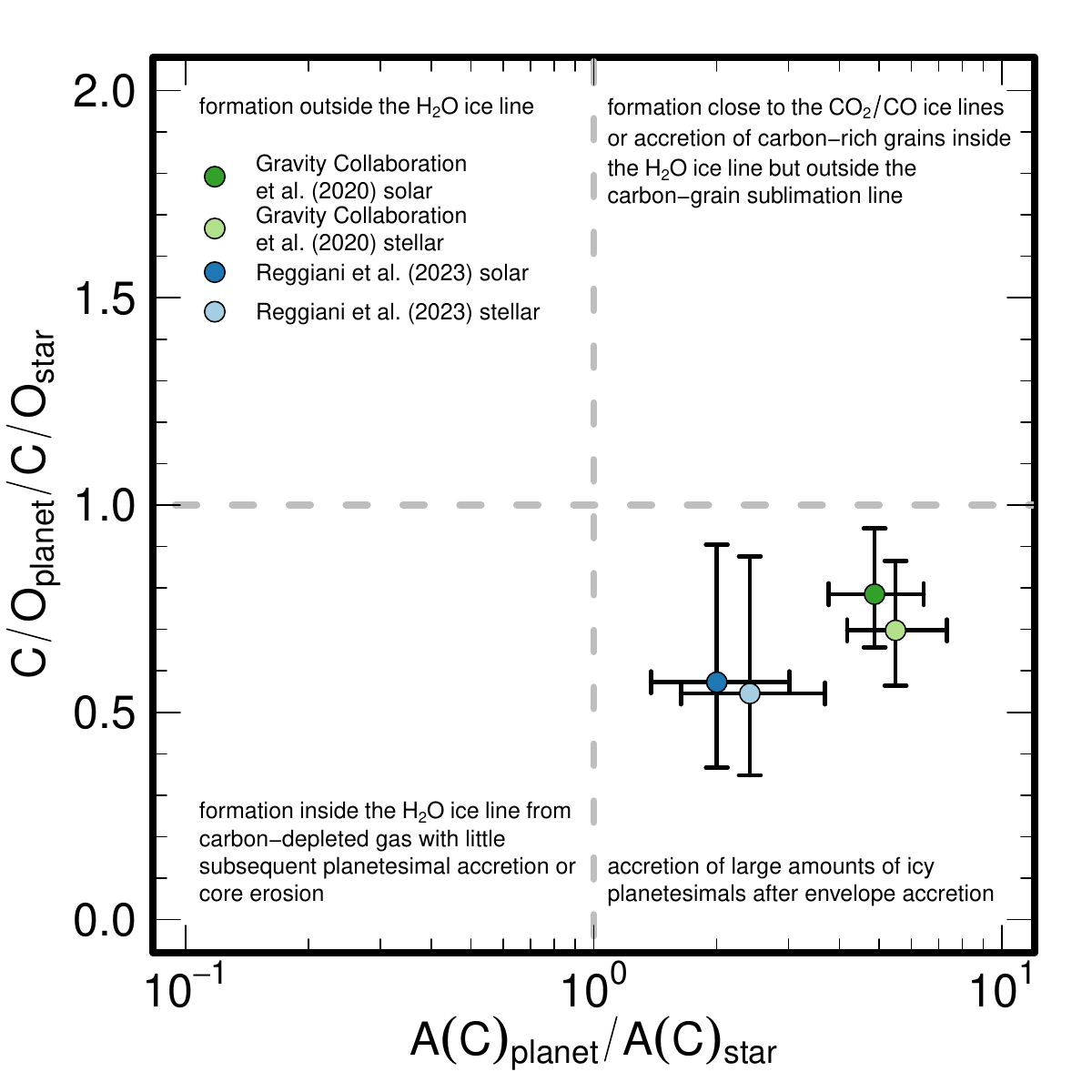}{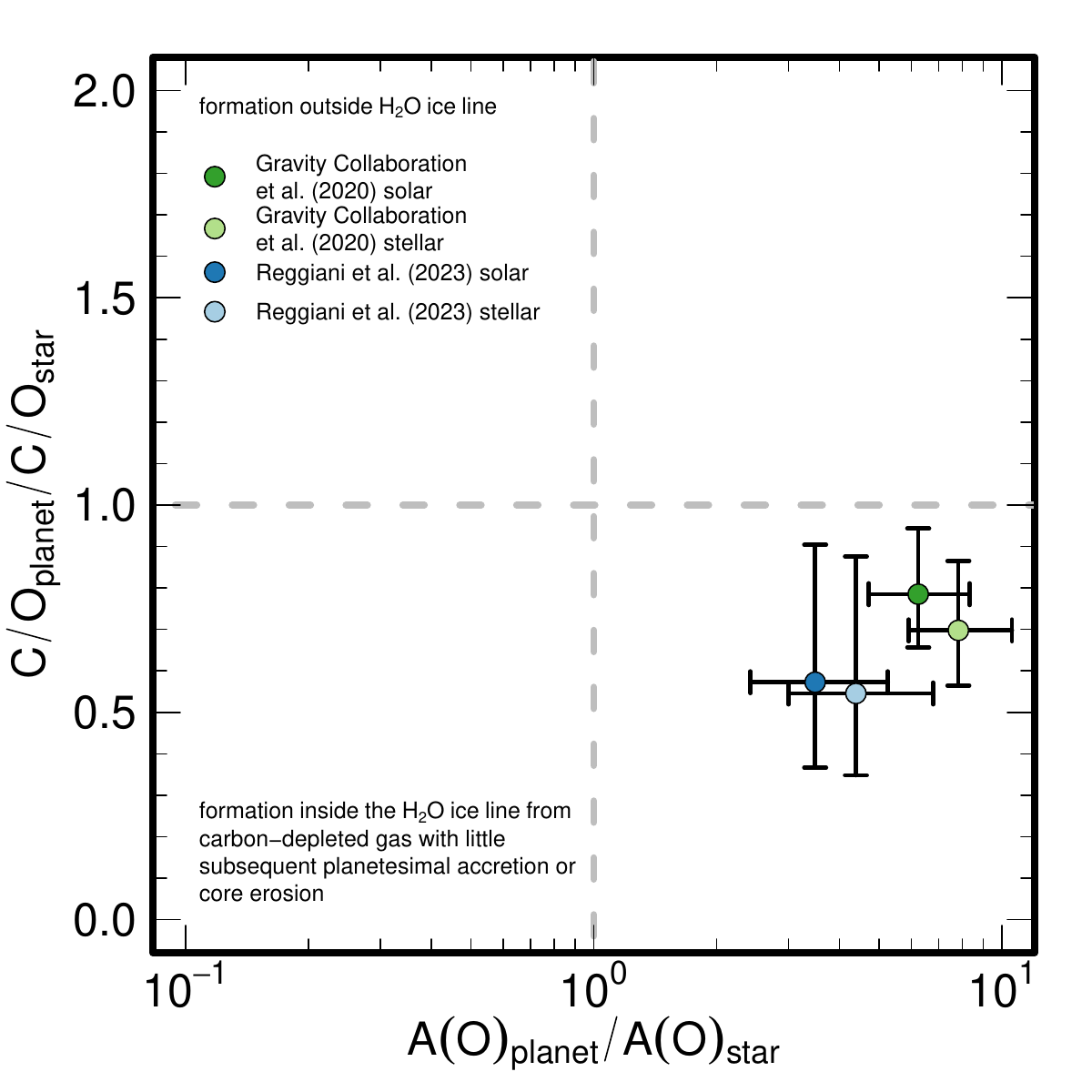}
\caption{On the left we present the planetary to stellar 
(protoplanetary disk proxy) C/O ratios of $\beta$ Pic b 
as a function of planet to stellar carbon abundances, and on the right we present the 
planetary to stellar C/O ratios as a function of planetary to stellar oxygen abundance. 
In dark and light green we present the planetary results of the \citet{gravity2020} study, 
scaled to solar and stellar (HD 181327) abundances, respectively. In dark and light blue we 
present the same, but for the carbon and oxygen abundances retrieved for $\beta$ Pic b from our study.
We see that regardless of the retrieval there is a difference in using solar and stellar abundances, 
portraying the importance of a dedicated stellar analysis. It is also clear that our retrieval 
has considerably changed the results. While there are visible differences, for $\beta$ Pic b 
our different analysis corroborate previsous conclusions about the planetary formation. It formed 
beyond the H$_2$O ice-line, accreting large amounts of icy planetesimals after envelope accretion.\label{co_star_planet}}
\end{figure*}

We present in Figure \ref{co_star_planet} the 
retrieved $(\textrm{C/O})_{\textrm{Planet}}/(\textrm{C/O})_{\textrm{Star}}$ as a function of 
$\textrm{A(C)}_{\textrm{Planet}}/\textrm{A(C)}_{\textrm{Star}}$ and as a function of 
$\textrm{A(O)}_{\textrm{Planet}}/\textrm{A(O)}_{\textrm{Star}}$, for both 
ours and \citet{gravity2020} retrievals. We present in both cases the abundances assuming solar C/O and 
carbon and oxygen for the host star, and our inference from HD 181327. In both panels we outline the 
different possible formation scenarios as described in the model presented by \citet{oberg2011}. 

As already pointed by \citet{gravity2020} we have little information about the 
relationship between the current orbit of $\beta$ Pic b and its formation 
location, and about possible variations of the locations of
the ice lines in systems such as that of $\beta$ Pic (no direct observations 
of the ice lines in $\beta$ Pic have been made to the best of our knowledge). 
However, based on the formation model presented by \cite{oberg2011}, 
our results indicate $\beta$ Pic b formed beyond the H$_2$O ice line by accreting large 
ammounts of icy planetesimals after envelope accretion. 

Even though we obtained different results in our planetary retrieval (see the metallicity difference), our stellar and planetary analysis corroborates the discussion presented by the Gravity Team. \citet{gravity2020} inferred $\beta$ Pic b's atmospheric C/O ratio is indicative $\beta$ Pic b slowly formed between the H$_2$O and CO$_2$ ice lines, via core-accretion. Its substellar C/O can be explained if the planet was enriched in oxygen by icy planetesimals. For more complete calculations we refer the reader to their study. We would also like to point out that even with the possible maximum systematic differences in the stellar non-LTE corrections from \citet{bergemann2021} (stellar C/O$=+0.43$) the conclusions drawn are the same. The planet would be located in the same quadrant as is in Figure \ref{co_star_planet}. We have, therefore, strong evidence on the formation scenario of this planet, as all results converge regardless of analysis (two different planetary atmospheric retrievals, and different assumptions to the composition of its protoplanetary disk pointing to the same conclusion).

\section{Conclusion}
\label{conclusion}

We find that the chemical pattern of HD 181327, a member of the $\beta$ Pic moving group, can be reliably used as a proxy to infer the chemical content of the molecular cloud form which $\beta$ Pic was formed. As such, and considering the abundances of $\beta$ Pic itself are not a reliable representation of its parent molecular cloud because of its intrinsic characteristics, we argue that the composition of the Hot Jupiter $\beta$ Pic b should be interpreted in light of HD 181327 composition.

We enumerate what we argue to be the most reliable ways to infer 
the composition of the protoplanetary disks from which exoplanets 
formed. 

We also performed our own retrieval of the parameters and abundances of $\beta$ Pic b. We retrieved a planetary metallicity that is fully in agreement to the abundance of HD 181327. We present carbon and oxygen abundances for both star and planet (stellar abundances are 3D non-LTE corrected). 
Our stellar C/O ratio is close to solar, but we found an important discrepancy in the abundances (in articular the metallicity) from our retrieval and the retrieval by \citet{gravity2020}. We attribute the differences to the our updated (recalibrated) data reduction pipeline. The combination of our stellar analysis and planetary retrieval indicate $\beta$ Pic. $b$ formed beyond the H$_2$O ice-line, and 
accreted large amounts of icy planetesimals after envelope accretion. It is important to point that our updated retrieval along with a new stellar analysis arrived at the same conclusion as previous studies, corroborating their conclusions and strongly reinforcing the proposed scenario for the formation of $\beta$ Pic $b$.

\section*{Acknowledgments}
We thank the referee for a careful read of our paper, and their comments that helped us improve the quality of our paper. 
The work of Henrique Reggiani was partially supported by NOIRLab, which is managed by the Association of Universities for Research in Astronomy (AURA) under a cooperative agreement with the National Science Foundation, and partially supported by a Carnegie Fellowship. Jhon Yana Galarza acknowledges support from their Carnegie Fellowship. 
This work made use of data collected with the Clay 6.5 meters Magellan 
Telescope. This work made use of ESO archival data, collected from its 
science portal. 
This work has made use of data from the European Space Agency (ESA)
mission {\it Gaia} (\url{https://www.cosmos.esa.int/gaia}), processed
by the {\it Gaia} Data Processing and Analysis Consortium (DPAC,
\url{https://www.cosmos.esa.int/web/gaia/dpac/consortium}). Funding for
the DPAC has been provided by national institutions, in particular the
institutions participating in the {\it Gaia} Multilateral Agreement.
This publication makes use of data products from the Two Micron All
Sky Survey, which is a joint project of the University of Massachusetts
and the Infrared Processing and Analysis Center/California Institute of
Technology, funded by the National Aeronautics and Space Administration
and the National Science Foundation.  This research has made use of the
NASA/IPAC Infrared Science Archive, which is funded by the National
Aeronautics and Space Administration and operated by the California
Institute of Technology. This research has made use of
NASA's Astrophysics Data System Bibliographic Services.  This research
has made use of the SIMBAD database, operated at CDS, Strasbourg,
France \citep{wenger2000}.  This research has made use of the VizieR
catalogue access tool, CDS, Strasbourg, France (DOI: 10.26093/cds/vizier).
The original description of the VizieR service was published in 2000,
A\&AS 143, 23 \citep{ochsenbein2000}.  This research has made use of the
NASA Exoplanet Archive, which is operated by the California Institute
of Technology, under contract with the National Aeronautics and Space
Administration under the Exoplanet Exploration Program.

\facilities{ LCO, CDS, Exoplanet Archive, Gaia, IRSA, Skymapper, Sloan,
TESS, ESO}

\software{\texttt{astropy} \citep{astropy2013,astropy2018},
          \texttt{Carpy} \citep{kelson2000,kelson2003,kelson2014},
	  \texttt{colte} \citep{casagrande2021},
          \texttt{IRAF} \citep{tody1986,tody1993},
          \texttt{isochrones} \citep{morton2015},
	  \texttt{lightkurve} \citep{lightkurve2018},
	  \texttt{MOOG} \citep{sneden1973},
          \texttt{MultiNest} \citep{feroz2008,feroz2009,feroz2019},
	  \texttt{numpy} \citep{harris2020},
          \texttt{pandas} \citep{mckinney2010,reback2020},
	  \texttt{PyMultinest} \citep{buchner2014},
	  \texttt{q$^{2}$} \citep{ramirez2014},
	  \texttt{RadVel} \citep{fulton2017,fulton2018},
          \texttt{scipy} \citep{virtanen2020}
          }
          


\bibliography{betapic}{}

\begin{thebibliography}{}
\expandafter\ifx\csname natexlab\endcsname\relax\def\natexlab#1{#1}\fi
\providecommand{\url}[1]{\href{#1}{#1}}
\providecommand{\dodoi}[1]{doi:~\href{http://doi.org/#1}{\nolinkurl{#1}}}
\providecommand{\doeprint}[1]{\href{http://ascl.net/#1}{\nolinkurl{http://ascl.net/#1}}}
\providecommand{\doarXiv}[1]{\href{https://arxiv.org/abs/#1}{\nolinkurl{https://arxiv.org/abs/#1}}}

\bibitem[{{Aerts} {et~al.}(2008){Aerts}, {Hekker}, {Desmet}, {Carrier}, {Zima},
  {Briquet}, \& {De Ridder}}]{aerts2008}
{Aerts}, C., {Hekker}, S., {Desmet}, M., {et~al.} 2008, in Precision
  Spectroscopy in Astrophysics, ed. N.~C. {Santos}, L.~{Pasquini}, A.~C.~M.
  {Correia}, \& M.~{Romaniello}, 161--164, \dodoi{10.1007/978-3-540-75485-5_35}

\bibitem[{{Ali-Dib}(2017)}]{ali-dib2017}
{Ali-Dib}, M. 2017, \mnras, 467, 2845, \dodoi{10.1093/mnras/stx260}

\bibitem[{{Ali-Dib} {et~al.}(2014){Ali-Dib}, {Mousis}, {Petit}, \&
  {Lunine}}]{ali-dib2014}
{Ali-Dib}, M., {Mousis}, O., {Petit}, J.-M., \& {Lunine}, J.~I. 2014, \apj,
  785, 125, \dodoi{10.1088/0004-637X/785/2/125}

\bibitem[{{Amarsi} \& {Asplund}(2017)}]{amarsi2017}
{Amarsi}, A.~M., \& {Asplund}, M. 2017, \mnras, 464, 264,
  \dodoi{10.1093/mnras/stw2445}

\bibitem[{{Amarsi} {et~al.}(2016){Amarsi}, {Lind}, {Asplund}, {Barklem}, \&
  {Collet}}]{amarsi2016}
{Amarsi}, A.~M., {Lind}, K., {Asplund}, M., {Barklem}, P.~S., \& {Collet}, R.
  2016, \mnras, 463, 1518, \dodoi{10.1093/mnras/stw2077}

\bibitem[{{Amarsi} {et~al.}(2019){Amarsi}, {Nissen}, \&
  {Sk{\'u}lad{\'o}ttir}}]{amarsi2019}
{Amarsi}, A.~M., {Nissen}, P.~E., \& {Sk{\'u}lad{\'o}ttir}, {\'A}. 2019, \aap,
  630, A104, \dodoi{10.1051/0004-6361/201936265}

\bibitem[{{Amarsi} {et~al.}(2020){Amarsi}, {Lind}, {Osorio}, {Nordlander},
  {Bergemann}, {Reggiani}, {Wang}, {Buder}, {Asplund}, {Barklem}, {Wehrhahn},
  {Sk{\'u}lad{\'o}ttir}, {Kobayashi}, {Karakas}, {Gao}, {Bland-Hawthorn}, {de
  Silva}, {Kos}, {Lewis}, {Martell}, {Sharma}, {Simpson}, {Zucker},
  {{\v{C}}otar}, {Horner}, \& {Galah Collaboration}}]{amarsi2020}
{Amarsi}, A.~M., {Lind}, K., {Osorio}, Y., {et~al.} 2020, \aap, 642, A62,
  \dodoi{10.1051/0004-6361/202038650}

\bibitem[{{Amundsen} {et~al.}(2014){Amundsen}, {Baraffe}, {Tremblin},
  {Manners}, {Hayek}, {Mayne}, \& {Acreman}}]{2014A&A...564A..59A}
{Amundsen}, D.~S., {Baraffe}, I., {Tremblin}, P., {et~al.} 2014, \aap, 564,
  A59, \dodoi{10.1051/0004-6361/201323169}

\bibitem[{{Andrews} {et~al.}(2019){Andrews}, {Anguiano}, {Chanam{\'e}},
  {Ag{\"u}eros}, {Lewis}, {Hayes}, \& {Majewski}}]{andrews2019}
{Andrews}, J.~J., {Anguiano}, B., {Chanam{\'e}}, J., {et~al.} 2019, \apj, 871,
  42, \dodoi{10.3847/1538-4357/aaf502}

\bibitem[{{Andrievsky} {et~al.}(2002){Andrievsky}, {Chernyshova}, {Paunzen},
  {Weiss}, {Korotin}, {Beletsky}, {Handler}, {Heiter}, {Korotina}, {St{\"u}tz},
  \& {Weber}}]{andrievsky2002}
{Andrievsky}, S.~M., {Chernyshova}, I.~V., {Paunzen}, E., {et~al.} 2002, \aap,
  396, 641, \dodoi{10.1051/0004-6361:20021423}

\bibitem[{{Arenou} {et~al.}(2018){Arenou}, {Luri}, {Babusiaux}, {Fabricius},
  {Helmi}, {Muraveva}, {Robin}, {Spoto}, {Vallenari}, {Antoja},
  {Cantat-Gaudin}, {Jordi}, {Leclerc}, {Reyl{\'e}}, {Romero-G{\'o}mez}, {Shih},
  {Soria}, {Barache}, {Bossini}, {Bragaglia}, {Breddels}, {Fabrizio},
  {Lambert}, {Marrese}, {Massari}, {Moitinho}, {Robichon}, {Ruiz-Dern},
  {Sordo}, {Veljanoski}, {Eyer}, {Jasniewicz}, {Pancino}, {Soubiran}, {Spagna},
  {Tanga}, {Turon}, \& {Zurbach}}]{arenou2018}
{Arenou}, F., {Luri}, X., {Babusiaux}, C., {et~al.} 2018, \aap, 616, A17,
  \dodoi{10.1051/0004-6361/201833234}

\bibitem[{{Asplund} {et~al.}(2021){Asplund}, {Amarsi}, \&
  {Grevesse}}]{asplund2021}
{Asplund}, M., {Amarsi}, A.~M., \& {Grevesse}, N. 2021, \aap, 653, A141,
  \dodoi{10.1051/0004-6361/202140445}

\bibitem[{{Asplund} {et~al.}(2009){Asplund}, {Grevesse}, {Sauval}, \&
  {Scott}}]{asplund2009}
{Asplund}, M., {Grevesse}, N., {Sauval}, A.~J., \& {Scott}, P. 2009, \araa, 47,
  481, \dodoi{10.1146/annurev.astro.46.060407.145222}

\bibitem[{{Astropy Collaboration} {et~al.}(2013){Astropy Collaboration},
  {Robitaille}, {Tollerud}, {Greenfield}, {Droettboom}, {Bray}, {Aldcroft},
  {Davis}, {Ginsburg}, {Price-Whelan}, {Kerzendorf}, {Conley}, {Crighton},
  {Barbary}, {Muna}, {Ferguson}, {Grollier}, {Parikh}, {Nair}, {Unther},
  {Deil}, {Woillez}, {Conseil}, {Kramer}, {Turner}, {Singer}, {Fox}, {Weaver},
  {Zabalza}, {Edwards}, {Azalee Bostroem}, {Burke}, {Casey}, {Crawford},
  {Dencheva}, {Ely}, {Jenness}, {Labrie}, {Lim}, {Pierfederici}, {Pontzen},
  {Ptak}, {Refsdal}, {Servillat}, \& {Streicher}}]{astropy2013}
{Astropy Collaboration}, {Robitaille}, T.~P., {Tollerud}, E.~J., {et~al.} 2013,
  \aap, 558, A33, \dodoi{10.1051/0004-6361/201322068}

\bibitem[{{Astropy Collaboration} {et~al.}(2018){Astropy Collaboration},
  {Price-Whelan}, {Sip{\H{o}}cz}, {G{\"u}nther}, {Lim}, {Crawford}, {Conseil},
  {Shupe}, {Craig}, {Dencheva}, {Ginsburg}, {VanderPlas}, {Bradley},
  {P{\'e}rez-Su{\'a}rez}, {de Val-Borro}, {Aldcroft}, {Cruz}, {Robitaille},
  {Tollerud}, {Ardelean}, {Babej}, {Bach}, {Bachetti}, {Bakanov}, {Bamford},
  {Barentsen}, {Barmby}, {Baumbach}, {Berry}, {Biscani}, {Boquien}, {Bostroem},
  {Bouma}, {Brammer}, {Bray}, {Breytenbach}, {Buddelmeijer}, {Burke},
  {Calderone}, {Cano Rodr{\'\i}guez}, {Cara}, {Cardoso}, {Cheedella}, {Copin},
  {Corrales}, {Crichton}, {D'Avella}, {Deil}, {Depagne}, {Dietrich}, {Donath},
  {Droettboom}, {Earl}, {Erben}, {Fabbro}, {Ferreira}, {Finethy}, {Fox},
  {Garrison}, {Gibbons}, {Goldstein}, {Gommers}, {Greco}, {Greenfield},
  {Groener}, {Grollier}, {Hagen}, {Hirst}, {Homeier}, {Horton}, {Hosseinzadeh},
  {Hu}, {Hunkeler}, {Ivezi{\'c}}, {Jain}, {Jenness}, {Kanarek}, {Kendrew},
  {Kern}, {Kerzendorf}, {Khvalko}, {King}, {Kirkby}, {Kulkarni}, {Kumar},
  {Lee}, {Lenz}, {Littlefair}, {Ma}, {Macleod}, {Mastropietro}, {McCully},
  {Montagnac}, {Morris}, {Mueller}, {Mumford}, {Muna}, {Murphy}, {Nelson},
  {Nguyen}, {Ninan}, {N{\"o}the}, {Ogaz}, {Oh}, {Parejko}, {Parley}, {Pascual},
  {Patil}, {Patil}, {Plunkett}, {Prochaska}, {Rastogi}, {Reddy Janga},
  {Sabater}, {Sakurikar}, {Seifert}, {Sherbert}, {Sherwood-Taylor}, {Shih},
  {Sick}, {Silbiger}, {Singanamalla}, {Singer}, {Sladen}, {Sooley},
  {Sornarajah}, {Streicher}, {Teuben}, {Thomas}, {Tremblay}, {Turner},
  {Terr{\'o}n}, {van Kerkwijk}, {de la Vega}, {Watkins}, {Weaver}, {Whitmore},
  {Woillez}, {Zabalza}, \& {Astropy Contributors}}]{astropy2018}
{Astropy Collaboration}, {Price-Whelan}, A.~M., {Sip{\H{o}}cz}, B.~M., {et~al.}
  2018, \aj, 156, 123, \dodoi{10.3847/1538-3881/aabc4f}

\bibitem[{{Bailer-Jones} {et~al.}(2021){Bailer-Jones}, {Rybizki}, {Fouesneau},
  {Demleitner}, \& {Andrae}}]{bailer-jones2021}
{Bailer-Jones}, C.~A.~L., {Rybizki}, J., {Fouesneau}, M., {Demleitner}, M., \&
  {Andrae}, R. 2021, \aj, 161, 147, \dodoi{10.3847/1538-3881/abd806}

\bibitem[{{Balmer} {et~al.}(2023){Balmer}, {Pueyo}, {Stolker}, {Reggiani},
  {Lacour}, {Maire}, {Molli{\`e}re}, {Nowak}, {Sing}, {Pourr{\'e}}, {Blunt},
  {Wang}, {Rickman}, {Henning}, {Ward-Duong}, {Abuter}, {Amorim},
  {Asensio-Torres}, {Benisty}, {Berger}, {Beust}, {Boccaletti}, {Bohn},
  {Bonnefoy}, {Bonnet}, {Bourdarot}, {Brandner}, {Cantalloube}, {Caselli},
  {Charnay}, {Chauvin}, {Chavez}, {Choquet}, {Christiaens}, {Cl{\'e}net},
  {Coud{\'e} du Foresto}, {Cridland}, {Dembet}, {Drescher}, {Duvert}, {Eckart},
  {Eisenhauer}, {Feuchtgruber}, {Garcia}, {Garcia Lopez}, {Gardner}, {Gendron},
  {Genzel}, {Gillessen}, {Girard}, {Haubois}, {Hei{\ss}el}, {Hinkley},
  {Hippler}, {Horrobin}, {Houll{\'e}}, {Hubert}, {Jocou}, {Kammerer},
  {Keppler}, {Kervella}, {Kreidberg}, {Lagrange}, {Lapeyr{\`e}re}, {Le
  Bouquin}, {L{\'e}na}, {Lutz}, {Mang}, {Marleau}, {M{\'e}rand}, {Monnier},
  {Mordasini}, {Mouillet}, {Nasedkin}, {Ott}, {Otten}, {Paladini}, {Paumard},
  {Perraut}, {Perrin}, {Pfuhl}, {Rameau}, {Rodet}, {Rustamkulov}, {Shangguan},
  {Shimizu}, {Straubmeier}, {Sturm}, {Tacconi}, {van Dishoeck}, {Vigan},
  {Vincent}, {Widmann}, {Wieprecht}, {Wiezorrek}, {Winterhalder}, {Woillez},
  {Yazici}, \& {Young}}]{balmer2023}
{Balmer}, W.~O., {Pueyo}, L., {Stolker}, T., {et~al.} 2023, arXiv e-prints,
  arXiv:2309.04403, \dodoi{10.48550/arXiv.2309.04403}

\bibitem[{{Bergemann} {et~al.}(2021){Bergemann}, {Hoppe}, {Semenova},
  {Carlsson}, {Yakovleva}, {Voronov}, {Bautista}, {Nemer}, {Belyaev},
  {Leenaarts}, {Mashonkina}, {Reiners}, \& {Ellwarth}}]{bergemann2021}
{Bergemann}, M., {Hoppe}, R., {Semenova}, E., {et~al.} 2021, \mnras, 508, 2236,
  \dodoi{10.1093/mnras/stab2160}

\bibitem[{{Bernstein} {et~al.}(2003){Bernstein}, {Shectman}, {Gunnels},
  {Mochnacki}, \& {Athey}}]{bernstein2003}
{Bernstein}, R., {Shectman}, S.~A., {Gunnels}, S.~M., {Mochnacki}, S., \&
  {Athey}, A.~E. 2003, in Society of Photo-Optical Instrumentation Engineers
  (SPIE) Conference Series, Vol. 4841, \procspie, ed. M.~{Iye} \& A.~F.~M.
  {Moorwood}, 1694--1704, \dodoi{10.1117/12.461502}

\bibitem[{{Bertran de Lis} {et~al.}(2016){Bertran de Lis}, {Allende Prieto},
  {Majewski}, {Schiavon}, {Holtzman}, {Shetrone}, {Carrera}, {Garc{\'\i}a
  P{\'e}rez}, {M{\'e}sz{\'a}ros}, {Frinchaboy}, {Hearty}, {Nidever},
  {Zasowski}, \& {Ge}}]{bertran2016}
{Bertran de Lis}, S., {Allende Prieto}, C., {Majewski}, S.~R., {et~al.} 2016,
  \aap, 590, A74, \dodoi{10.1051/0004-6361/201527827}

\bibitem[{{Beust} \& {Morbidelli}(2000)}]{beust2000}
{Beust}, H., \& {Morbidelli}, A. 2000, \icarus, 143, 170,
  \dodoi{10.1006/icar.1999.6238}

\bibitem[{{Booth} {et~al.}(2017){Booth}, {Clarke}, {Madhusudhan}, \&
  {Ilee}}]{booth2017}
{Booth}, R.~A., {Clarke}, C.~J., {Madhusudhan}, N., \& {Ilee}, J.~D. 2017,
  \mnras, 469, 3994, \dodoi{10.1093/mnras/stx1103}

\bibitem[{{Bovy}(2016)}]{bovy2016}
{Bovy}, J. 2016, \apj, 817, 49, \dodoi{10.3847/0004-637X/817/1/49}

\bibitem[{{Buchner} {et~al.}(2014){Buchner}, {Georgakakis}, {Nandra}, {Hsu},
  {Rangel}, {Brightman}, {Merloni}, {Salvato}, {Donley}, \&
  {Kocevski}}]{buchner2014}
{Buchner}, J., {Georgakakis}, A., {Nandra}, K., {et~al.} 2014, \aap, 564, A125,
  \dodoi{10.1051/0004-6361/201322971}

\bibitem[{{Buder} {et~al.}(2021){Buder}, {Sharma}, {Kos}, {Amarsi},
  {Nordlander}, {Lind}, {Martell}, {Asplund}, {Bland-Hawthorn}, {Casey}, {de
  Silva}, {D'Orazi}, {Freeman}, {Hayden}, {Lewis}, {Lin}, {Schlesinger},
  {Simpson}, {Stello}, {Zucker}, {Zwitter}, {Beeson}, {Buck}, {Casagrande},
  {Clark}, {{\v{C}}otar}, {da Costa}, {de Grijs}, {Feuillet}, {Horner},
  {Kafle}, {Khanna}, {Kobayashi}, {Liu}, {Montet}, {Nandakumar}, {Nataf},
  {Ness}, {Spina}, {Tepper-Garc{\'\i}a}, {Ting}, {Traven},
  {Vogrin{\v{c}}i{\v{c}}}, {Wittenmyer}, {Wyse}, {{\v{Z}}erjal}, \& {GALAH
  Collaboration}}]{buder2021}
{Buder}, S., {Sharma}, S., {Kos}, J., {et~al.} 2021, \mnras, 506, 150,
  \dodoi{10.1093/mnras/stab1242}

\bibitem[{{Caffau} {et~al.}(2011){Caffau}, {Ludwig}, {Steffen}, {Freytag}, \&
  {Bonifacio}}]{Caffau2011SoPh..268..255C}
{Caffau}, E., {Ludwig}, H.~G., {Steffen}, M., {Freytag}, B., \& {Bonifacio}, P.
  2011, \solphys, 268, 255, \dodoi{10.1007/s11207-010-9541-4}

\bibitem[{{C{\'a}novas} {et~al.}(2019){C{\'a}novas}, {Cantero}, {Cieza},
  {Bombrun}, {Lammers}, {Mer{\'\i}n}, {Mora}, {Ribas}, \&
  {Ru{\'\i}z-Rodr{\'\i}guez}}]{canovas2019}
{C{\'a}novas}, H., {Cantero}, C., {Cieza}, L., {et~al.} 2019, \aap, 626, A80,
  \dodoi{10.1051/0004-6361/201935321}

\bibitem[{{Capitanio} {et~al.}(2017){Capitanio}, {Lallement}, {Vergely},
  {Elyajouri}, \& {Monreal-Ibero}}]{capitanio2017}
{Capitanio}, L., {Lallement}, R., {Vergely}, J.~L., {Elyajouri}, M., \&
  {Monreal-Ibero}, A. 2017, \aap, 606, A65, \dodoi{10.1051/0004-6361/201730831}

\bibitem[{{Casagrande} {et~al.}(2021){Casagrande}, {Lin}, {Rains}, {Liu},
  {Buder}, {Horner}, {Asplund}, {Lewis}, {Martell}, {Nordlander}, {Stello},
  {Ting}, {Wittenmyer}, {Bland-Hawthorn}, {Casey}, {De Silva}, {D'Orazi},
  {Freeman}, {Hayden}, {Kos}, {Lind}, {Schlesinger}, {Sharma}, {Simpson},
  {Zucker}, \& {Zwitter}}]{casagrande2021}
{Casagrande}, L., {Lin}, J., {Rains}, A.~D., {et~al.} 2021, \mnras, 507, 2684,
  \dodoi{10.1093/mnras/stab2304}

\bibitem[{{Casey}(2014)}]{casey2014}
{Casey}, A.~R. 2014, PhD thesis, Australian National University, Canberra

\bibitem[{{Castelli} \& {Kurucz}(2003)}]{Castelli2003}
{Castelli}, F., \& {Kurucz}, R.~L. 2003, in Modelling of Stellar Atmospheres,
  ed. N.~{Piskunov}, W.~W. {Weiss}, \& D.~F. {Gray}, Vol. 210, A20,
  \dodoi{10.48550/arXiv.astro-ph/0405087}

\bibitem[{{Chilcote} {et~al.}(2017){Chilcote}, {Pueyo}, {De Rosa}, {Vargas},
  {Macintosh}, {Bailey}, {Barman}, {Bauman}, {Bruzzone}, {Bulger}, {Burrows},
  {Cardwell}, {Chen}, {Cotten}, {Dillon}, {Doyon}, {Draper}, {Duch{\^e}ne},
  {Dunn}, {Erikson}, {Fitzgerald}, {Follette}, {Gavel}, {Goodsell}, {Graham},
  {Greenbaum}, {Hartung}, {Hibon}, {Hung}, {Ingraham}, {Kalas}, {Konopacky},
  {Larkin}, {Maire}, {Marchis}, {Marley}, {Marois}, {Metchev},
  {Millar-Blanchaer}, {Morzinski}, {Nielsen}, {Norton}, {Oppenheimer},
  {Palmer}, {Patience}, {Perrin}, {Poyneer}, {Rajan}, {Rameau},
  {Rantakyr{\"o}}, {Sadakuni}, {Saddlemyer}, {Savransky}, {Schneider}, {Serio},
  {Sivaramakrishnan}, {Song}, {Soummer}, {Thomas}, {Wallace}, {Wang},
  {Ward-Duong}, {Wiktorowicz}, \& {Wolff}}]{chilcote2017}
{Chilcote}, J., {Pueyo}, L., {De Rosa}, R.~J., {et~al.} 2017, \aj, 153, 182,
  \dodoi{10.3847/1538-3881/aa63e9}

\bibitem[{{Choi} {et~al.}(2016){Choi}, {Dotter}, {Conroy}, {Cantiello},
  {Paxton}, \& {Johnson}}]{choi2016}
{Choi}, J., {Dotter}, A., {Conroy}, C., {et~al.} 2016, \apj, 823, 102,
  \dodoi{10.3847/0004-637X/823/2/102}

\bibitem[{{Cridland} {et~al.}(2019{\natexlab{a}}){Cridland}, {Eistrup}, \& {van
  Dishoeck}}]{cridland2019a}
{Cridland}, A.~J., {Eistrup}, C., \& {van Dishoeck}, E.~F. 2019{\natexlab{a}},
  \aap, 627, A127, \dodoi{10.1051/0004-6361/201834378}

\bibitem[{{Cridland} {et~al.}(2016){Cridland}, {Pudritz}, \&
  {Alessi}}]{cridland2016}
{Cridland}, A.~J., {Pudritz}, R.~E., \& {Alessi}, M. 2016, \mnras, 461, 3274,
  \dodoi{10.1093/mnras/stw1511}

\bibitem[{{Cridland} {et~al.}(2019{\natexlab{b}}){Cridland}, {van Dishoeck},
  {Alessi}, \& {Pudritz}}]{cridland2019b}
{Cridland}, A.~J., {van Dishoeck}, E.~F., {Alessi}, M., \& {Pudritz}, R.~E.
  2019{\natexlab{b}}, \aap, 632, A63, \dodoi{10.1051/0004-6361/201936105}

\bibitem[{{Cridland} {et~al.}(2020){Cridland}, {van Dishoeck}, {Alessi}, \&
  {Pudritz}}]{cridland2020}
---. 2020, \aap, 642, A229, \dodoi{10.1051/0004-6361/202038767}

\bibitem[{{Crifo} {et~al.}(1997){Crifo}, {Vidal-Madjar}, {Lallement}, {Ferlet},
  \& {Gerbaldi}}]{crifo1997}
{Crifo}, F., {Vidal-Madjar}, A., {Lallement}, R., {Ferlet}, R., \& {Gerbaldi},
  M. 1997, in ESA Special Publication, Vol. 402, Hipparcos - Venice '97, ed.
  R.~M. {Bonnet}, E.~{H{\o}g}, P.~L. {Bernacca}, L.~{Emiliani}, A.~{Blaauw},
  C.~{Turon}, J.~{Kovalevsky}, L.~{Lindegren}, H.~{Hassan}, M.~{Bouffard},
  B.~{Strim}, D.~{Heger}, M.~A.~C. {Perryman}, \& L.~{Woltjer}, 437--440

\bibitem[{{Crozet} {et~al.}(2023){Crozet}, {Morin}, {Ross}, {Bellotti},
  {Donati}, {Fouqu{\'e}}, {Moutou}, {Petit}, {Carmona}, {K{\'o}sp{\'a}l},
  {Adam}, \& {Tokaryk}}]{crozet2023}
{Crozet}, P., {Morin}, J., {Ross}, A.~J., {et~al.} 2023, arXiv e-prints,
  arXiv:2310.04497, \dodoi{10.48550/arXiv.2310.04497}

\bibitem[{{De Rosa} {et~al.}(2020){De Rosa}, {Esposito}, {Gibbs}, {Bailey},
  {Fitzgerald}, {Chilcote}, {Duch{\^e}ne}, {Konopacky}, {Macintosh},
  {Millar-Blanchaer}, {Nguyen}, {Nielsen}, {Perrin}, {Rameau}, \&
  {Wang}}]{DeRosa2020SPIE11447E..5AD}
{De Rosa}, R.~J., {Esposito}, T.~M., {Gibbs}, A., {et~al.} 2020, in Society of
  Photo-Optical Instrumentation Engineers (SPIE) Conference Series, Vol. 11447,
  Society of Photo-Optical Instrumentation Engineers (SPIE) Conference Series,
  114475A, \dodoi{10.1117/12.2561071}

\bibitem[{{De Silva} {et~al.}(2006{\natexlab{a}}){De Silva}, {Sneden},
  {Paulson}, {Asplund}, {Bland-Hawthorn}, {Bessell}, \&
  {Freeman}}]{desilva2006}
{De Silva}, G.~M., {Sneden}, C., {Paulson}, D.~B., {et~al.} 2006{\natexlab{a}},
  \aj, 131, 455, \dodoi{10.1086/497968}

\bibitem[{{De Silva} {et~al.}(2006{\natexlab{b}}){De Silva}, {Sneden},
  {Paulson}, {Asplund}, {Bland-Hawthorn}, {Bessell}, \&
  {Freeman}}]{desilva2011}
---. 2006{\natexlab{b}}, \aj, 131, 455, \dodoi{10.1086/497968}

\bibitem[{{Dotter}(2016)}]{dotter2016}
{Dotter}, A. 2016, \apjs, 222, 8, \dodoi{10.3847/0067-0049/222/1/8}

\bibitem[{{Drummond} {et~al.}(2019){Drummond}, {Carter}, {H{\'e}brard},
  {Mayne}, {Sing}, {Evans}, \& {Goyal}}]{2019MNRAS.486.1123D}
{Drummond}, B., {Carter}, A.~L., {H{\'e}brard}, E., {et~al.} 2019, \mnras, 486,
  1123, \dodoi{10.1093/mnras/stz909}

\bibitem[{{Drummond} {et~al.}(2016){Drummond}, {Tremblin}, {Baraffe},
  {Amundsen}, {Mayne}, {Venot}, \& {Goyal}}]{2016A&A...594A..69D}
{Drummond}, B., {Tremblin}, P., {Baraffe}, I., {et~al.} 2016, \aap, 594, A69,
  \dodoi{10.1051/0004-6361/201628799}

\bibitem[{{Eistrup} {et~al.}(2018){Eistrup}, {Walsh}, \& {van
  Dishoeck}}]{eistrup2018}
{Eistrup}, C., {Walsh}, C., \& {van Dishoeck}, E.~F. 2018, \aap, 613, A14,
  \dodoi{10.1051/0004-6361/201731302}

\bibitem[{{Espinoza} {et~al.}(2017){Espinoza}, {Fortney}, {Miguel},
  {Thorngren}, \& {Murray-Clay}}]{espinoza2017}
{Espinoza}, N., {Fortney}, J.~J., {Miguel}, Y., {Thorngren}, D., \&
  {Murray-Clay}, R. 2017, \apjl, 838, L9, \dodoi{10.3847/2041-8213/aa65ca}

\bibitem[{{Evans} {et~al.}(2018){Evans}, {Riello}, {De Angeli}, {Carrasco},
  {Montegriffo}, {Fabricius}, {Jordi}, {Palaversa}, {Diener}, {Busso},
  {Cacciari}, {van Leeuwen}, {Burgess}, {Davidson}, {Harrison}, {Hodgkin},
  {Pancino}, {Richards}, {Altavilla}, {Balaguer-N{\'u}{\~n}ez}, {Barstow},
  {Bellazzini}, {Brown}, {Castellani}, {Cocozza}, {De Luise}, {Delgado},
  {Ducourant}, {Galleti}, {Gilmore}, {Giuffrida}, {Holl}, {Kewley}, {Koposov},
  {Marinoni}, {Marrese}, {Osborne}, {Piersimoni}, {Portell}, {Pulone},
  {Ragaini}, {Sanna}, {Terrett}, {Walton}, {Wevers}, \&
  {Wyrzykowski}}]{evans2018}
{Evans}, D.~W., {Riello}, M., {De Angeli}, F., {et~al.} 2018, \aap, 616, A4,
  \dodoi{10.1051/0004-6361/201832756}

\bibitem[{{Fabricius} {et~al.}(2021){Fabricius}, {Luri}, {Arenou}, {Babusiaux},
  {Helmi}, {Muraveva}, {Reyl{\'e}}, {Spoto}, {Vallenari}, {Antoja}, {Balbinot},
  {Barache}, {Bauchet}, {Bragaglia}, {Busonero}, {Cantat-Gaudin}, {Carrasco},
  {Diakit{\'e}}, {Fabrizio}, {Figueras}, {Garcia-Gutierrez}, {Garofalo},
  {Jordi}, {Kervella}, {Khanna}, {Leclerc}, {Licata}, {Lambert}, {Marrese},
  {Masip}, {Ramos}, {Robichon}, {Robin}, {Romero-G{\'o}mez}, {Rubele}, \&
  {Weiler}}]{fabricius2021}
{Fabricius}, C., {Luri}, X., {Arenou}, F., {et~al.} 2021, \aap, 649, A5,
  \dodoi{10.1051/0004-6361/202039834}

\bibitem[{{Feinstein} {et~al.}(2019){Feinstein}, {Montet}, {Foreman-Mackey},
  {Bedell}, {Saunders}, {Bean}, {Christiansen}, {Hedges}, {Luger}, {Scolnic},
  \& {Cardoso}}]{feinstein2019}
{Feinstein}, A.~D., {Montet}, B.~T., {Foreman-Mackey}, D., {et~al.} 2019,
  \pasp, 131, 094502, \dodoi{10.1088/1538-3873/ab291c}

\bibitem[{{Feng} {et~al.}(2022){Feng}, {Butler}, {Vogt}, {Clement}, {Tinney},
  {Cui}, {Aizawa}, {Jones}, {Bailey}, {Burt}, {Carter}, {Crane}, {Flammini
  Dotti}, {Holden}, {Ma}, {Ogihara}, {Oppenheimer}, {O'Toole}, {Shectman},
  {Wittenmyer}, {Wang}, {Wright}, \& {Xuan}}]{2022ApJS..262...21F}
{Feng}, F., {Butler}, R.~P., {Vogt}, S.~S., {et~al.} 2022, \apjs, 262, 21,
  \dodoi{10.3847/1538-4365/ac7e57}

\bibitem[{{Ferlet} {et~al.}(1987){Ferlet}, {Hobbs}, \&
  {Vidal-Madjar}}]{ferlet1987}
{Ferlet}, R., {Hobbs}, L.~M., \& {Vidal-Madjar}, A. 1987, \aap, 185, 267

\bibitem[{{Feroz} \& {Hobson}(2008)}]{feroz2008}
{Feroz}, F., \& {Hobson}, M.~P. 2008, \mnras, 384, 449,
  \dodoi{10.1111/j.1365-2966.2007.12353.x}

\bibitem[{{Feroz} {et~al.}(2009){Feroz}, {Hobson}, \& {Bridges}}]{feroz2009}
{Feroz}, F., {Hobson}, M.~P., \& {Bridges}, M. 2009, \mnras, 398, 1601,
  \dodoi{10.1111/j.1365-2966.2009.14548.x}

\bibitem[{{Feroz} {et~al.}(2019){Feroz}, {Hobson}, {Cameron}, \&
  {Pettitt}}]{feroz2019}
{Feroz}, F., {Hobson}, M.~P., {Cameron}, E., \& {Pettitt}, A.~N. 2019, The Open
  Journal of Astrophysics, 2, 10, \dodoi{10.21105/astro.1306.2144}

\bibitem[{{Folsom} {et~al.}(2012){Folsom}, {Bagnulo}, {Wade}, {Alecian},
  {Landstreet}, {Marsden}, \& {Waite}}]{folsom2012}
{Folsom}, C.~P., {Bagnulo}, S., {Wade}, G.~A., {et~al.} 2012, \mnras, 422,
  2072, \dodoi{10.1111/j.1365-2966.2012.20718.x}

\bibitem[{{Fulton} {et~al.}(2017){Fulton}, {Petigura}, {Blunt}, \&
  {Sinukoff}}]{fulton2017}
{Fulton}, B.~J., {Petigura}, E.~A., {Blunt}, S., \& {Sinukoff}, E. 2017,
  {RadVel: The Radial Velocity Fitting Toolkit}, \dodoi{10.5281/zenodo.580821}

\bibitem[{{Fulton} {et~al.}(2018){Fulton}, {Petigura}, {Blunt}, \&
  {Sinukoff}}]{fulton2018}
---. 2018, \pasp, 130, 044504, \dodoi{10.1088/1538-3873/aaaaa8}

\bibitem[{{Gaia Collaboration} {et~al.}(2016){Gaia Collaboration}, {Prusti},
  {de Bruijne}, {Brown}, {Vallenari}, {Babusiaux}, {Bailer-Jones}, {Bastian},
  {Biermann}, {Evans}, {Eyer}, {Jansen}, {Jordi}, {Klioner}, {Lammers},
  {Lindegren}, {Luri}, {Mignard}, {Milligan}, {Panem}, {Poinsignon},
  {Pourbaix}, {Randich}, {Sarri}, {Sartoretti}, {Siddiqui}, {Soubiran},
  {Valette}, {van Leeuwen}, {Walton}, {Aerts}, {Arenou}, {Cropper}, {Drimmel},
  {H{\o}g}, {Katz}, {Lattanzi}, {O'Mullane}, {Grebel}, {Holland}, {Huc},
  {Passot}, {Bramante}, {Cacciari}, {Casta{\~n}eda}, {Chaoul}, {Cheek}, {De
  Angeli}, {Fabricius}, {Guerra}, {Hern{\'a}ndez}, {Jean-Antoine-Piccolo},
  {Masana}, {Messineo}, {Mowlavi}, {Nienartowicz}, {Ord{\'o}{\~n}ez-Blanco},
  {Panuzzo}, {Portell}, {Richards}, {Riello}, {Seabroke}, {Tanga},
  {Th{\'e}venin}, {Torra}, {Els}, {Gracia-Abril}, {Comoretto},
  {Garcia-Reinaldos}, {Lock}, {Mercier}, {Altmann}, {Andrae}, {Astraatmadja},
  {Bellas-Velidis}, {Benson}, {Berthier}, {Blomme}, {Busso}, {Carry},
  {Cellino}, {Clementini}, {Cowell}, {Creevey}, {Cuypers}, {Davidson}, {De
  Ridder}, {de Torres}, {Delchambre}, {Dell'Oro}, {Ducourant}, {Fr{\'e}mat},
  {Garc{\'\i}a-Torres}, {Gosset}, {Halbwachs}, {Hambly}, {Harrison}, {Hauser},
  {Hestroffer}, {Hodgkin}, {Huckle}, {Hutton}, {Jasniewicz}, {Jordan},
  {Kontizas}, {Korn}, {Lanzafame}, {Manteiga}, {Moitinho}, {Muinonen},
  {Osinde}, {Pancino}, {Pauwels}, {Petit}, {Recio-Blanco}, {Robin}, {Sarro},
  {Siopis}, {Smith}, {Smith}, {Sozzetti}, {Thuillot}, {van Reeven}, {Viala},
  {Abbas}, {Abreu Aramburu}, {Accart}, {Aguado}, {Allan}, {Allasia},
  {Altavilla}, {{\'A}lvarez}, {Alves}, {Anderson}, {Andrei}, {Anglada Varela},
  {Antiche}, {Antoja}, {Ant{\'o}n}, {Arcay}, {Atzei}, {Ayache}, {Bach},
  {Baker}, {Balaguer-N{\'u}{\~n}ez}, {Barache}, {Barata}, {Barbier}, {Barblan},
  {Baroni}, {Barrado y Navascu{\'e}s}, {Barros}, {Barstow}, {Becciani},
  {Bellazzini}, {Bellei}, {Bello Garc{\'\i}a}, {Belokurov}, {Bendjoya},
  {Berihuete}, {Bianchi}, {Bienaym{\'e}}, {Billebaud}, {Blagorodnova},
  {Blanco-Cuaresma}, {Boch}, {Bombrun}, {Borrachero}, {Bouquillon}, {Bourda},
  {Bouy}, {Bragaglia}, {Breddels}, {Brouillet}, {Br{\"u}semeister},
  {Bucciarelli}, {Budnik}, {Burgess}, {Burgon}, {Burlacu}, {Busonero}, {Buzzi},
  {Caffau}, {Cambras}, {Campbell}, {Cancelliere}, {Cantat-Gaudin}, {Carlucci},
  {Carrasco}, {Castellani}, {Charlot}, {Charnas}, {Charvet}, {Chassat},
  {Chiavassa}, {Clotet}, {Cocozza}, {Collins}, {Collins}, {Costigan}, {Crifo},
  {Cross}, {Crosta}, {Crowley}, {Dafonte}, {Damerdji}, {Dapergolas}, {David},
  {David}, {De Cat}, {de Felice}, {de Laverny}, {De Luise}, {De March}, {de
  Martino}, {de Souza}, {Debosscher}, {del Pozo}, {Delbo}, {Delgado},
  {Delgado}, {di Marco}, {Di Matteo}, {Diakite}, {Distefano}, {Dolding}, {Dos
  Anjos}, {Drazinos}, {Dur{\'a}n}, {Dzigan}, {Ecale}, {Edvardsson}, {Enke},
  {Erdmann}, {Escolar}, {Espina}, {Evans}, {Eynard Bontemps}, {Fabre},
  {Fabrizio}, {Faigler}, {Falc{\~a}o}, {Farr{\`a}s Casas}, {Faye}, {Federici},
  {Fedorets}, {Fern{\'a}ndez-Hern{\'a}ndez}, {Fernique}, {Fienga}, {Figueras},
  {Filippi}, {Findeisen}, {Fonti}, {Fouesneau}, {Fraile}, {Fraser}, {Fuchs},
  {Furnell}, {Gai}, {Galleti}, {Galluccio}, {Garabato}, {Garc{\'\i}a-Sedano},
  {Gar{\'e}}, {Garofalo}, {Garralda}, {Gavras}, {Gerssen}, {Geyer}, {Gilmore},
  {Girona}, {Giuffrida}, {Gomes}, {Gonz{\'a}lez-Marcos},
  {Gonz{\'a}lez-N{\'u}{\~n}ez}, {Gonz{\'a}lez-Vidal}, {Granvik}, {Guerrier},
  {Guillout}, {Guiraud}, {G{\'u}rpide}, {Guti{\'e}rrez-S{\'a}nchez}, {Guy},
  {Haigron}, {Hatzidimitriou}, {Haywood}, {Heiter}, {Helmi}, {Hobbs},
  {Hofmann}, {Holl}, {Holland}, {Hunt}, {Hypki}, {Icardi}, {Irwin}, {Jevardat
  de Fombelle}, {Jofr{\'e}}, {Jonker}, {Jorissen}, {Julbe}, {Karampelas},
  {Kochoska}, {Kohley}, {Kolenberg}, {Kontizas}, {Koposov}, {Kordopatis},
  {Koubsky}, {Kowalczyk}, {Krone-Martins}, {Kudryashova}, {Kull}, {Bachchan},
  {Lacoste-Seris}, {Lanza}, {Lavigne}, {Le Poncin-Lafitte}, {Lebreton},
  {Lebzelter}, {Leccia}, {Leclerc}, {Lecoeur-Taibi}, {Lemaitre}, {Lenhardt},
  {Leroux}, {Liao}, {Licata}, {Lindstr{\o}m}, {Lister}, {Livanou}, {Lobel},
  {L{\"o}ffler}, {L{\'o}pez}, {Lopez-Lozano}, {Lorenz}, {Loureiro},
  {MacDonald}, {Magalh{\~a}es Fernandes}, {Managau}, {Mann}, {Mantelet},
  {Marchal}, {Marchant}, {Marconi}, {Marie}, {Marinoni}, {Marrese},
  {Marschalk{\'o}}, {Marshall}, {Mart{\'\i}n-Fleitas}, {Martino}, {Mary},
  {Matijevi{\v{c}}}, {Mazeh}, {McMillan}, {Messina}, {Mestre}, {Michalik},
  {Millar}, {Miranda}, {Molina}, {Molinaro}, {Molinaro}, {Moln{\'a}r},
  {Moniez}, {Montegriffo}, {Monteiro}, {Mor}, {Mora}, {Morbidelli}, {Morel},
  {Morgenthaler}, {Morley}, {Morris}, {Mulone}, {Muraveva}, {Musella},
  {Narbonne}, {Nelemans}, {Nicastro}, {Noval}, {Ord{\'e}novic},
  {Ordieres-Mer{\'e}}, {Osborne}, {Pagani}, {Pagano}, {Pailler}, {Palacin},
  {Palaversa}, {Parsons}, {Paulsen}, {Pecoraro}, {Pedrosa}, {Pentik{\"a}inen},
  {Pereira}, {Pichon}, {Piersimoni}, {Pineau}, {Plachy}, {Plum}, {Poujoulet},
  {Pr{\v{s}}a}, {Pulone}, {Ragaini}, {Rago}, {Rambaux}, {Ramos-Lerate},
  {Ranalli}, {Rauw}, {Read}, {Regibo}, {Renk}, {Reyl{\'e}}, {Ribeiro},
  {Rimoldini}, {Ripepi}, {Riva}, {Rixon}, {Roelens}, {Romero-G{\'o}mez},
  {Rowell}, {Royer}, {Rudolph}, {Ruiz-Dern}, {Sadowski}, {Sagrist{\`a}
  Sell{\'e}s}, {Sahlmann}, {Salgado}, {Salguero}, {Sarasso}, {Savietto},
  {Schnorhk}, {Schultheis}, {Sciacca}, {Segol}, {Segovia}, {Segransan},
  {Serpell}, {Shih}, {Smareglia}, {Smart}, {Smith}, {Solano}, {Solitro},
  {Sordo}, {Soria Nieto}, {Souchay}, {Spagna}, {Spoto}, {Stampa}, {Steele},
  {Steidelm{\"u}ller}, {Stephenson}, {Stoev}, {Suess}, {S{\"u}veges}, {Surdej},
  {Szabados}, {Szegedi-Elek}, {Tapiador}, {Taris}, {Tauran}, {Taylor},
  {Teixeira}, {Terrett}, {Tingley}, {Trager}, {Turon}, {Ulla}, {Utrilla},
  {Valentini}, {van Elteren}, {Van Hemelryck}, {van Leeuwen}, {Varadi},
  {Vecchiato}, {Veljanoski}, {Via}, {Vicente}, {Vogt}, {Voss}, {Votruba},
  {Voutsinas}, {Walmsley}, {Weiler}, {Weingrill}, {Werner}, {Wevers},
  {Whitehead}, {Wyrzykowski}, {Yoldas}, {{\v{Z}}erjal}, {Zucker}, {Zurbach},
  {Zwitter}, {Alecu}, {Allen}, {Allende Prieto}, {Amorim},
  {Anglada-Escud{\'e}}, {Arsenijevic}, {Azaz}, {Balm}, {Beck}, {Bernstein},
  {Bigot}, {Bijaoui}, {Blasco}, {Bonfigli}, {Bono}, {Boudreault}, {Bressan},
  {Brown}, {Brunet}, {Bunclark}, {Buonanno}, {Butkevich}, {Carret}, {Carrion},
  {Chemin}, {Ch{\'e}reau}, {Corcione}, {Darmigny}, {de Boer}, {de Teodoro}, {de
  Zeeuw}, {Delle Luche}, {Domingues}, {Dubath}, {Fodor}, {Fr{\'e}zouls},
  {Fries}, {Fustes}, {Fyfe}, {Gallardo}, {Gallegos}, {Gardiol}, {Gebran},
  {Gomboc}, {G{\'o}mez}, {Grux}, {Gueguen}, {Heyrovsky}, {Hoar}, {Iannicola},
  {Isasi Parache}, {Janotto}, {Joliet}, {Jonckheere}, {Keil}, {Kim},
  {Klagyivik}, {Klar}, {Knude}, {Kochukhov}, {Kolka}, {Kos}, {Kutka}, {Lainey},
  {LeBouquin}, {Liu}, {Loreggia}, {Makarov}, {Marseille}, {Martayan},
  {Martinez-Rubi}, {Massart}, {Meynadier}, {Mignot}, {Munari}, {Nguyen},
  {Nordlander}, {Ocvirk}, {O'Flaherty}, {Olias Sanz}, {Ortiz}, {Osorio},
  {Oszkiewicz}, {Ouzounis}, {Palmer}, {Park}, {Pasquato}, {Peltzer}, {Peralta},
  {P{\'e}turaud}, {Pieniluoma}, {Pigozzi}, {Poels}, {Prat}, {Prod'homme},
  {Raison}, {Rebordao}, {Risquez}, {Rocca-Volmerange}, {Rosen}, {Ruiz-Fuertes},
  {Russo}, {Sembay}, {Serraller Vizcaino}, {Short}, {Siebert}, {Silva},
  {Sinachopoulos}, {Slezak}, {Soffel}, {Sosnowska}, {Strai{\v{z}}ys}, {ter
  Linden}, {Terrell}, {Theil}, {Tiede}, {Troisi}, {Tsalmantza}, {Tur},
  {Vaccari}, {Vachier}, {Valles}, {Van Hamme}, {Veltz}, {Virtanen}, {Wallut},
  {Wichmann}, {Wilkinson}, {Ziaeepour}, \& {Zschocke}}]{gaia2016}
{Gaia Collaboration}, {Prusti}, T., {de Bruijne}, J.~H.~J., {et~al.} 2016,
  \aap, 595, A1, \dodoi{10.1051/0004-6361/201629272}

\bibitem[{{Gaia Collaboration} {et~al.}(2018){Gaia Collaboration}, {Brown},
  {Vallenari}, {Prusti}, {de Bruijne}, {Babusiaux}, {Bailer-Jones}, {Biermann},
  {Evans}, {Eyer}, {Jansen}, {Jordi}, {Klioner}, {Lammers}, {Lindegren},
  {Luri}, {Mignard}, {Panem}, {Pourbaix}, {Randich}, {Sartoretti}, {Siddiqui},
  {Soubiran}, {van Leeuwen}, {Walton}, {Arenou}, {Bastian}, {Cropper},
  {Drimmel}, {Katz}, {Lattanzi}, {Bakker}, {Cacciari}, {Casta{\~n}eda},
  {Chaoul}, {Cheek}, {De Angeli}, {Fabricius}, {Guerra}, {Holl}, {Masana},
  {Messineo}, {Mowlavi}, {Nienartowicz}, {Panuzzo}, {Portell}, {Riello},
  {Seabroke}, {Tanga}, {Th{\'e}venin}, {Gracia-Abril}, {Comoretto},
  {Garcia-Reinaldos}, {Teyssier}, {Altmann}, {Andrae}, {Audard},
  {Bellas-Velidis}, {Benson}, {Berthier}, {Blomme}, {Burgess}, {Busso},
  {Carry}, {Cellino}, {Clementini}, {Clotet}, {Creevey}, {Davidson}, {De
  Ridder}, {Delchambre}, {Dell'Oro}, {Ducourant},
  {Fern{\'a}ndez-Hern{\'a}ndez}, {Fouesneau}, {Fr{\'e}mat}, {Galluccio},
  {Garc{\'\i}a-Torres}, {Gonz{\'a}lez-N{\'u}{\~n}ez}, {Gonz{\'a}lez-Vidal},
  {Gosset}, {Guy}, {Halbwachs}, {Hambly}, {Harrison}, {Hern{\'a}ndez},
  {Hestroffer}, {Hodgkin}, {Hutton}, {Jasniewicz}, {Jean-Antoine-Piccolo},
  {Jordan}, {Korn}, {Krone-Martins}, {Lanzafame}, {Lebzelter}, {L{\"o}ffler},
  {Manteiga}, {Marrese}, {Mart{\'\i}n-Fleitas}, {Moitinho}, {Mora}, {Muinonen},
  {Osinde}, {Pancino}, {Pauwels}, {Petit}, {Recio-Blanco}, {Richards},
  {Rimoldini}, {Robin}, {Sarro}, {Siopis}, {Smith}, {Sozzetti}, {S{\"u}veges},
  {Torra}, {van Reeven}, {Abbas}, {Abreu Aramburu}, {Accart}, {Aerts},
  {Altavilla}, {{\'A}lvarez}, {Alvarez}, {Alves}, {Anderson}, {Andrei},
  {Anglada Varela}, {Antiche}, {Antoja}, {Arcay}, {Astraatmadja}, {Bach},
  {Baker}, {Balaguer-N{\'u}{\~n}ez}, {Balm}, {Barache}, {Barata}, {Barbato},
  {Barblan}, {Barklem}, {Barrado}, {Barros}, {Barstow}, {Bartholom{\'e}
  Mu{\~n}oz}, {Bassilana}, {Becciani}, {Bellazzini}, {Berihuete}, {Bertone},
  {Bianchi}, {Bienaym{\'e}}, {Blanco-Cuaresma}, {Boch}, {Boeche}, {Bombrun},
  {Borrachero}, {Bossini}, {Bouquillon}, {Bourda}, {Bragaglia}, {Bramante},
  {Breddels}, {Bressan}, {Brouillet}, {Br{\"u}semeister}, {Brugaletta},
  {Bucciarelli}, {Burlacu}, {Busonero}, {Butkevich}, {Buzzi}, {Caffau},
  {Cancelliere}, {Cannizzaro}, {Cantat-Gaudin}, {Carballo}, {Carlucci},
  {Carrasco}, {Casamiquela}, {Castellani}, {Castro-Ginard}, {Charlot},
  {Chemin}, {Chiavassa}, {Cocozza}, {Costigan}, {Cowell}, {Crifo}, {Crosta},
  {Crowley}, {Cuypers}, {Dafonte}, {Damerdji}, {Dapergolas}, {David}, {David},
  {de Laverny}, {De Luise}, {De March}, {de Martino}, {de Souza}, {de Torres},
  {Debosscher}, {del Pozo}, {Delbo}, {Delgado}, {Delgado}, {Di Matteo},
  {Diakite}, {Diener}, {Distefano}, {Dolding}, {Drazinos}, {Dur{\'a}n},
  {Edvardsson}, {Enke}, {Eriksson}, {Esquej}, {Eynard Bontemps}, {Fabre},
  {Fabrizio}, {Faigler}, {Falc{\~a}o}, {Farr{\`a}s Casas}, {Federici},
  {Fedorets}, {Fernique}, {Figueras}, {Filippi}, {Findeisen}, {Fonti},
  {Fraile}, {Fraser}, {Fr{\'e}zouls}, {Gai}, {Galleti}, {Garabato},
  {Garc{\'\i}a-Sedano}, {Garofalo}, {Garralda}, {Gavel}, {Gavras}, {Gerssen},
  {Geyer}, {Giacobbe}, {Gilmore}, {Girona}, {Giuffrida}, {Glass}, {Gomes},
  {Granvik}, {Gueguen}, {Guerrier}, {Guiraud}, {Guti{\'e}rrez-S{\'a}nchez},
  {Haigron}, {Hatzidimitriou}, {Hauser}, {Haywood}, {Heiter}, {Helmi}, {Heu},
  {Hilger}, {Hobbs}, {Hofmann}, {Holland}, {Huckle}, {Hypki}, {Icardi},
  {Jan{\ss}en}, {Jevardat de Fombelle}, {Jonker}, {Juh{\'a}sz}, {Julbe},
  {Karampelas}, {Kewley}, {Klar}, {Kochoska}, {Kohley}, {Kolenberg},
  {Kontizas}, {Kontizas}, {Koposov}, {Kordopatis}, {Kostrzewa-Rutkowska},
  {Koubsky}, {Lambert}, {Lanza}, {Lasne}, {Lavigne}, {Le Fustec}, {Le
  Poncin-Lafitte}, {Lebreton}, {Leccia}, {Leclerc}, {Lecoeur-Taibi},
  {Lenhardt}, {Leroux}, {Liao}, {Licata}, {Lindstr{\o}m}, {Lister}, {Livanou},
  {Lobel}, {L{\'o}pez}, {Managau}, {Mann}, {Mantelet}, {Marchal}, {Marchant},
  {Marconi}, {Marinoni}, {Marschalk{\'o}}, {Marshall}, {Martino}, {Marton},
  {Mary}, {Massari}, {Matijevi{\v{c}}}, {Mazeh}, {McMillan}, {Messina},
  {Michalik}, {Millar}, {Molina}, {Molinaro}, {Moln{\'a}r}, {Montegriffo},
  {Mor}, {Morbidelli}, {Morel}, {Morris}, {Mulone}, {Muraveva}, {Musella},
  {Nelemans}, {Nicastro}, {Noval}, {O'Mullane}, {Ord{\'e}novic},
  {Ord{\'o}{\~n}ez-Blanco}, {Osborne}, {Pagani}, {Pagano}, {Pailler},
  {Palacin}, {Palaversa}, {Panahi}, {Pawlak}, {Piersimoni}, {Pineau}, {Plachy},
  {Plum}, {Poggio}, {Poujoulet}, {Pr{\v{s}}a}, {Pulone}, {Racero}, {Ragaini},
  {Rambaux}, {Ramos-Lerate}, {Regibo}, {Reyl{\'e}}, {Riclet}, {Ripepi}, {Riva},
  {Rivard}, {Rixon}, {Roegiers}, {Roelens}, {Romero-G{\'o}mez}, {Rowell},
  {Royer}, {Ruiz-Dern}, {Sadowski}, {Sagrist{\`a} Sell{\'e}s}, {Sahlmann},
  {Salgado}, {Salguero}, {Sanna}, {Santana-Ros}, {Sarasso}, {Savietto},
  {Schultheis}, {Sciacca}, {Segol}, {Segovia}, {S{\'e}gransan}, {Shih},
  {Siltala}, {Silva}, {Smart}, {Smith}, {Solano}, {Solitro}, {Sordo}, {Soria
  Nieto}, {Souchay}, {Spagna}, {Spoto}, {Stampa}, {Steele},
  {Steidelm{\"u}ller}, {Stephenson}, {Stoev}, {Suess}, {Surdej}, {Szabados},
  {Szegedi-Elek}, {Tapiador}, {Taris}, {Tauran}, {Taylor}, {Teixeira},
  {Terrett}, {Teyssandier}, {Thuillot}, {Titarenko}, {Torra Clotet}, {Turon},
  {Ulla}, {Utrilla}, {Uzzi}, {Vaillant}, {Valentini}, {Valette}, {van Elteren},
  {Van Hemelryck}, {van Leeuwen}, {Vaschetto}, {Vecchiato}, {Veljanoski},
  {Viala}, {Vicente}, {Vogt}, {von Essen}, {Voss}, {Votruba}, {Voutsinas},
  {Walmsley}, {Weiler}, {Wertz}, {Wevers}, {Wyrzykowski}, {Yoldas},
  {{\v{Z}}erjal}, {Ziaeepour}, {Zorec}, {Zschocke}, {Zucker}, {Zurbach}, \&
  {Zwitter}}]{gaia2018}
{Gaia Collaboration}, {Brown}, A.~G.~A., {Vallenari}, A., {et~al.} 2018, \aap,
  616, A1, \dodoi{10.1051/0004-6361/201833051}

\bibitem[{{Gaia Collaboration} {et~al.}(2021){Gaia Collaboration}, {Brown},
  {Vallenari}, {Prusti}, {de Bruijne}, {Babusiaux}, {Biermann}, {Creevey},
  {Evans}, {Eyer}, {Hutton}, {Jansen}, {Jordi}, {Klioner}, {Lammers},
  {Lindegren}, {Luri}, {Mignard}, {Panem}, {Pourbaix}, {Randich}, {Sartoretti},
  {Soubiran}, {Walton}, {Arenou}, {Bailer-Jones}, {Bastian}, {Cropper},
  {Drimmel}, {Katz}, {Lattanzi}, {van Leeuwen}, {Bakker}, {Cacciari},
  {Casta{\~n}eda}, {De Angeli}, {Ducourant}, {Fabricius}, {Fouesneau},
  {Fr{\'e}mat}, {Guerra}, {Guerrier}, {Guiraud}, {Jean-Antoine Piccolo},
  {Masana}, {Messineo}, {Mowlavi}, {Nicolas}, {Nienartowicz}, {Pailler},
  {Panuzzo}, {Riclet}, {Roux}, {Seabroke}, {Sordo}, {Tanga}, {Th{\'e}venin},
  {Gracia-Abril}, {Portell}, {Teyssier}, {Altmann}, {Andrae}, {Bellas-Velidis},
  {Benson}, {Berthier}, {Blomme}, {Brugaletta}, {Burgess}, {Busso}, {Carry},
  {Cellino}, {Cheek}, {Clementini}, {Damerdji}, {Davidson}, {Delchambre},
  {Dell'Oro}, {Fern{\'a}ndez-Hern{\'a}ndez}, {Galluccio}, {Garc{\'\i}a-Lario},
  {Garcia-Reinaldos}, {Gonz{\'a}lez-N{\'u}{\~n}ez}, {Gosset}, {Haigron},
  {Halbwachs}, {Hambly}, {Harrison}, {Hatzidimitriou}, {Heiter},
  {Hern{\'a}ndez}, {Hestroffer}, {Hodgkin}, {Holl}, {Jan{\ss}en}, {Jevardat de
  Fombelle}, {Jordan}, {Krone-Martins}, {Lanzafame}, {L{\"o}ffler}, {Lorca},
  {Manteiga}, {Marchal}, {Marrese}, {Moitinho}, {Mora}, {Muinonen}, {Osborne},
  {Pancino}, {Pauwels}, {Petit}, {Recio-Blanco}, {Richards}, {Riello},
  {Rimoldini}, {Robin}, {Roegiers}, {Rybizki}, {Sarro}, {Siopis}, {Smith},
  {Sozzetti}, {Ulla}, {Utrilla}, {van Leeuwen}, {van Reeven}, {Abbas}, {Abreu
  Aramburu}, {Accart}, {Aerts}, {Aguado}, {Ajaj}, {Altavilla}, {{\'A}lvarez},
  {{\'A}lvarez Cid-Fuentes}, {Alves}, {Anderson}, {Anglada Varela}, {Antoja},
  {Audard}, {Baines}, {Baker}, {Balaguer-N{\'u}{\~n}ez}, {Balbinot}, {Balog},
  {Barache}, {Barbato}, {Barros}, {Barstow}, {Bartolom{\'e}}, {Bassilana},
  {Bauchet}, {Baudesson-Stella}, {Becciani}, {Bellazzini}, {Bernet}, {Bertone},
  {Bianchi}, {Blanco-Cuaresma}, {Boch}, {Bombrun}, {Bossini}, {Bouquillon},
  {Bragaglia}, {Bramante}, {Breedt}, {Bressan}, {Brouillet}, {Bucciarelli},
  {Burlacu}, {Busonero}, {Butkevich}, {Buzzi}, {Caffau}, {Cancelliere},
  {C{\'a}novas}, {Cantat-Gaudin}, {Carballo}, {Carlucci}, {Carnerero},
  {Carrasco}, {Casamiquela}, {Castellani}, {Castro-Ginard}, {Castro Sampol},
  {Chaoul}, {Charlot}, {Chemin}, {Chiavassa}, {Cioni}, {Comoretto}, {Cooper},
  {Cornez}, {Cowell}, {Crifo}, {Crosta}, {Crowley}, {Dafonte}, {Dapergolas},
  {David}, {David}, {de Laverny}, {De Luise}, {De March}, {De Ridder}, {de
  Souza}, {de Teodoro}, {de Torres}, {del Peloso}, {del Pozo}, {Delbo},
  {Delgado}, {Delgado}, {Delisle}, {Di Matteo}, {Diakite}, {Diener},
  {Distefano}, {Dolding}, {Eappachen}, {Edvardsson}, {Enke}, {Esquej}, {Fabre},
  {Fabrizio}, {Faigler}, {Fedorets}, {Fernique}, {Fienga}, {Figueras},
  {Fouron}, {Fragkoudi}, {Fraile}, {Franke}, {Gai}, {Garabato},
  {Garcia-Gutierrez}, {Garc{\'\i}a-Torres}, {Garofalo}, {Gavras}, {Gerlach},
  {Geyer}, {Giacobbe}, {Gilmore}, {Girona}, {Giuffrida}, {Gomel}, {Gomez},
  {Gonzalez-Santamaria}, {Gonz{\'a}lez-Vidal}, {Granvik},
  {Guti{\'e}rrez-S{\'a}nchez}, {Guy}, {Hauser}, {Haywood}, {Helmi}, {Hidalgo},
  {Hilger}, {H{\l}adczuk}, {Hobbs}, {Holland}, {Huckle}, {Jasniewicz},
  {Jonker}, {Juaristi Campillo}, {Julbe}, {Karbevska}, {Kervella}, {Khanna},
  {Kochoska}, {Kontizas}, {Kordopatis}, {Korn}, {Kostrzewa-Rutkowska},
  {Kruszy{\'n}ska}, {Lambert}, {Lanza}, {Lasne}, {Le Campion}, {Le Fustec},
  {Lebreton}, {Lebzelter}, {Leccia}, {Leclerc}, {Lecoeur-Taibi}, {Liao},
  {Licata}, {Lindstr{\o}m}, {Lister}, {Livanou}, {Lobel}, {Madrero Pardo},
  {Managau}, {Mann}, {Marchant}, {Marconi}, {Marcos Santos}, {Marinoni},
  {Marocco}, {Marshall}, {Martin Polo}, {Mart{\'\i}n-Fleitas}, {Masip},
  {Massari}, {Mastrobuono-Battisti}, {Mazeh}, {McMillan}, {Messina},
  {Michalik}, {Millar}, {Mints}, {Molina}, {Molinaro}, {Moln{\'a}r},
  {Montegriffo}, {Mor}, {Morbidelli}, {Morel}, {Morris}, {Mulone}, {Munoz},
  {Muraveva}, {Murphy}, {Musella}, {Noval}, {Ord{\'e}novic}, {Orr{\`u}},
  {Osinde}, {Pagani}, {Pagano}, {Palaversa}, {Palicio}, {Panahi}, {Pawlak},
  {Pe{\~n}alosa Esteller}, {Penttil{\"a}}, {Piersimoni}, {Pineau}, {Plachy},
  {Plum}, {Poggio}, {Poretti}, {Poujoulet}, {Pr{\v{s}}a}, {Pulone}, {Racero},
  {Ragaini}, {Rainer}, {Raiteri}, {Rambaux}, {Ramos}, {Ramos-Lerate}, {Re
  Fiorentin}, {Regibo}, {Reyl{\'e}}, {Ripepi}, {Riva}, {Rixon}, {Robichon},
  {Robin}, {Roelens}, {Rohrbasser}, {Romero-G{\'o}mez}, {Rowell}, {Royer},
  {Rybicki}, {Sadowski}, {Sagrist{\`a} Sell{\'e}s}, {Sahlmann}, {Salgado},
  {Salguero}, {Samaras}, {Sanchez Gimenez}, {Sanna}, {Santove{\~n}a},
  {Sarasso}, {Schultheis}, {Sciacca}, {Segol}, {Segovia}, {S{\'e}gransan},
  {Semeux}, {Shahaf}, {Siddiqui}, {Siebert}, {Siltala}, {Slezak}, {Smart},
  {Solano}, {Solitro}, {Souami}, {Souchay}, {Spagna}, {Spoto}, {Steele},
  {Steidelm{\"u}ller}, {Stephenson}, {S{\"u}veges}, {Szabados}, {Szegedi-Elek},
  {Taris}, {Tauran}, {Taylor}, {Teixeira}, {Thuillot}, {Tonello}, {Torra},
  {Torra}, {Turon}, {Unger}, {Vaillant}, {van Dillen}, {Vanel}, {Vecchiato},
  {Viala}, {Vicente}, {Voutsinas}, {Weiler}, {Wevers}, {Wyrzykowski}, {Yoldas},
  {Yvard}, {Zhao}, {Zorec}, {Zucker}, {Zurbach}, \& {Zwitter}}]{gaia2021}
---. 2021, \aap, 649, A1, \dodoi{10.1051/0004-6361/202039657}

\bibitem[{{Galland} {et~al.}(2006){Galland}, {Lagrange}, {Udry}, {Chelli},
  {Pepe}, {Beuzit}, \& {Mayor}}]{galland2006}
{Galland}, F., {Lagrange}, A.~M., {Udry}, S., {et~al.} 2006, \aap, 447, 355,
  \dodoi{10.1051/0004-6361:20054080}

\bibitem[{{Galli} {et~al.}(2020){Galli}, {Bouy}, {Olivares}, {Miret-Roig},
  {Sarro}, {Barrado}, {Berihuete}, \& {Brandner}}]{galli2020}
{Galli}, P.~A.~B., {Bouy}, H., {Olivares}, J., {et~al.} 2020, \aap, 634, A98,
  \dodoi{10.1051/0004-6361/201936708}

\bibitem[{{Galli} {et~al.}(2019){Galli}, {Loinard}, {Bouy}, {Sarro},
  {Ortiz-Le{\'o}n}, {Dzib}, {Olivares}, {Heyer}, {Hernandez},
  {Rom{\'a}n-Z{\'u}{\~n}iga}, {Kounkel}, \& {Covey}}]{galli2019}
{Galli}, P.~A.~B., {Loinard}, L., {Bouy}, H., {et~al.} 2019, \aap, 630, A137,
  \dodoi{10.1051/0004-6361/201935928}

\bibitem[{{Garc{\'\i}a P{\'e}rez} {et~al.}(2016){Garc{\'\i}a P{\'e}rez},
  {Allende Prieto}, {Holtzman}, {Shetrone}, {M{\'e}sz{\'a}ros}, {Bizyaev},
  {Carrera}, {Cunha}, {Garc{\'\i}a-Hern{\'a}ndez}, {Johnson}, {Majewski},
  {Nidever}, {Schiavon}, {Shane}, {Smith}, {Sobeck}, {Troup}, {Zamora},
  {Weinberg}, {Bovy}, {Eisenstein}, {Feuillet}, {Frinchaboy}, {Hayden},
  {Hearty}, {Nguyen}, {O'Connell}, {Pinsonneault}, {Wilson}, \&
  {Zasowski}}]{garcia-perez2016}
{Garc{\'\i}a P{\'e}rez}, A.~E., {Allende Prieto}, C., {Holtzman}, J.~A.,
  {et~al.} 2016, \aj, 151, 144, \dodoi{10.3847/0004-6256/151/6/144}

\bibitem[{{Goyal} {et~al.}(2018){Goyal}, {Mayne}, {Sing}, {Drummond},
  {Tremblin}, {Amundsen}, {Evans}, {Carter}, {Spake}, {Baraffe}, {Nikolov},
  {Manners}, {Chabrier}, \& {Hebrard}}]{2018MNRAS.474.5158G}
{Goyal}, J.~M., {Mayne}, N., {Sing}, D.~K., {et~al.} 2018, \mnras, 474, 5158,
  \dodoi{10.1093/mnras/stx3015}

\bibitem[{{GRAVITY Collaboration} {et~al.}(2020){GRAVITY Collaboration},
  {Nowak}, {Lacour}, {Molli{\`e}re}, {Wang}, {Charnay}, {van Dishoeck},
  {Abuter}, {Amorim}, {Berger}, {Beust}, {Bonnefoy}, {Bonnet}, {Brandner},
  {Buron}, {Cantalloube}, {Collin}, {Chapron}, {Cl{\'e}net}, {Coud{\'e} Du
  Foresto}, {de Zeeuw}, {Dembet}, {Dexter}, {Duvert}, {Eckart}, {Eisenhauer},
  {F{\"o}rster Schreiber}, {F{\'e}dou}, {Garcia Lopez}, {Gao}, {Gendron},
  {Genzel}, {Gillessen}, {Hau{\ss}mann}, {Henning}, {Hippler}, {Hubert},
  {Jocou}, {Kervella}, {Lagrange}, {Lapeyr{\`e}re}, {Le Bouquin}, {L{\'e}na},
  {Maire}, {Ott}, {Paumard}, {Paladini}, {Perraut}, {Perrin}, {Pueyo}, {Pfuhl},
  {Rabien}, {Rau}, {Rodr{\'\i}guez-Coira}, {Rousset}, {Scheithauer},
  {Shangguan}, {Straub}, {Straubmeier}, {Sturm}, {Tacconi}, {Vincent},
  {Widmann}, {Wieprecht}, {Wiezorrek}, {Woillez}, {Yazici}, \&
  {Ziegler}}]{gravity2020}
{GRAVITY Collaboration}, {Nowak}, M., {Lacour}, S., {et~al.} 2020, \aap, 633,
  A110, \dodoi{10.1051/0004-6361/201936898}

\bibitem[{{Guillot}(2010)}]{Guillot2010A&A...520A..27G}
{Guillot}, T. 2010, \aap, 520, A27, \dodoi{10.1051/0004-6361/200913396}

\bibitem[{{Hambly} {et~al.}(2018){Hambly}, {Cropper}, {Boudreault}, {Crowley},
  {Kohley}, {de Bruijne}, {Dolding}, {Fabricius}, {Seabroke}, {Davidson},
  {Rowell}, {Collins}, {Cross}, {Mart{\'\i}n-Fleitas}, {Baker}, {Smith},
  {Sartoretti}, {Marchal}, {Katz}, {De Angeli}, {Busso}, {Riello}, {Allende
  Prieto}, {Els}, {Corcione}, {Masana}, {Luri}, {Chassat}, {Fusero},
  {Pasquier}, {V{\'e}tel}, {Sarri}, \& {Gare}}]{hambly2018}
{Hambly}, N.~C., {Cropper}, M., {Boudreault}, S., {et~al.} 2018, \aap, 616,
  A15, \dodoi{10.1051/0004-6361/201832716}

\bibitem[{Harris {et~al.}(2020)Harris, Millman, van~der Walt, Gommers,
  Virtanen, Cournapeau, Wieser, Taylor, Berg, Smith, Kern, Picus, Hoyer, van
  Kerkwijk, Brett, Haldane, del R{\'{i}}o, Wiebe, Peterson,
  G{\'{e}}rard-Marchant, Sheppard, Reddy, Weckesser, Abbasi, Gohlke, \&
  Oliphant}]{harris2020}
Harris, C.~R., Millman, K.~J., van~der Walt, S.~J., {et~al.} 2020, Nature, 585,
  357, \dodoi{10.1038/s41586-020-2649-2}

\bibitem[{{Harsono} {et~al.}(2015){Harsono}, {Bruderer}, \& {van
  Dishoeck}}]{harsono2015}
{Harsono}, D., {Bruderer}, S., \& {van Dishoeck}, E.~F. 2015, \aap, 582, A41,
  \dodoi{10.1051/0004-6361/201525966}

\bibitem[{{Hawkins} {et~al.}(2020){Hawkins}, {Lucey}, {Ting}, {Ji}, {Katzberg},
  {Thompson}, {El-Badry}, {Teske}, {Nelson}, \& {Carrillo}}]{hawkins2020}
{Hawkins}, K., {Lucey}, M., {Ting}, Y.-S., {et~al.} 2020, \mnras, 492, 1164,
  \dodoi{10.1093/mnras/stz3132}

\bibitem[{{Healy} \& {McCullough}(2020)}]{healy2020}
{Healy}, B.~F., \& {McCullough}, P.~R. 2020, \apj, 903, 99,
  \dodoi{10.3847/1538-4357/abbc03}

\bibitem[{{Healy} {et~al.}(2021){Healy}, {McCullough}, \&
  {Schlaufman}}]{healy2021}
{Healy}, B.~F., {McCullough}, P.~R., \& {Schlaufman}, K.~C. 2021, \apj, 923,
  23, \dodoi{10.3847/1538-4357/ac281d}

\bibitem[{{Healy} {et~al.}(2023){Healy}, {McCullough}, {Schlaufman}, \&
  {Kovacs}}]{healy2023}
{Healy}, B.~F., {McCullough}, P.~R., {Schlaufman}, K.~C., \& {Kovacs}, G. 2023,
  \apj, 944, 39, \dodoi{10.3847/1538-4357/acad7b}

\bibitem[{{Heiter}(2002)}]{heiter2002}
{Heiter}, U. 2002, \aap, 381, 959, \dodoi{10.1051/0004-6361:20011593}

\bibitem[{{H{\o}g} {et~al.}(2000){H{\o}g}, {Fabricius}, {Makarov}, {Urban},
  {Corbin}, {Wycoff}, {Bastian}, {Schwekendiek}, \& {Wicenec}}]{hog2000}
{H{\o}g}, E., {Fabricius}, C., {Makarov}, V.~V., {et~al.} 2000, \aap, 355, L27

\bibitem[{{Kamp} {et~al.}(2001){Kamp}, {Iliev}, {Paunzen}, {Pintado}, {Solano},
  \& {Barzova}}]{kamp2001}
{Kamp}, I., {Iliev}, I.~K., {Paunzen}, E., {et~al.} 2001, \aap, 375, 899,
  \dodoi{10.1051/0004-6361:20010886}

\bibitem[{{Kelson}(2003)}]{kelson2003}
{Kelson}, D.~D. 2003, \pasp, 115, 688, \dodoi{10.1086/375502}

\bibitem[{{Kelson} {et~al.}(2000){Kelson}, {Illingworth}, {van Dokkum}, \&
  {Franx}}]{kelson2000}
{Kelson}, D.~D., {Illingworth}, G.~D., {van Dokkum}, P.~G., \& {Franx}, M.
  2000, \apj, 531, 159, \dodoi{10.1086/308445}

\bibitem[{{Kelson} {et~al.}(2014){Kelson}, {Williams}, {Dressler}, {McCarthy},
  {Shectman}, {Mulchaey}, {Villanueva}, {Crane}, \& {Quadri}}]{kelson2014}
{Kelson}, D.~D., {Williams}, R.~J., {Dressler}, A., {et~al.} 2014, \apj, 783,
  110, \dodoi{10.1088/0004-637X/783/2/110}

\bibitem[{{Kobayashi} {et~al.}(2020){Kobayashi}, {Karakas}, \&
  {Lugaro}}]{kobayashi2020}
{Kobayashi}, C., {Karakas}, A.~I., \& {Lugaro}, M. 2020, \apj, 900, 179,
  \dodoi{10.3847/1538-4357/abae65}

\bibitem[{{Koen}(2003)}]{koen2003a}
{Koen}, C. 2003, \mnras, 341, 1385, \dodoi{10.1046/j.1365-8711.2003.06509.x}

\bibitem[{{Koen} {et~al.}(2003){Koen}, {Balona}, {Khadaroo}, {Lane},
  {Prinsloo}, {Smith}, \& {Laney}}]{koen2003b}
{Koen}, C., {Balona}, L.~A., {Khadaroo}, K., {et~al.} 2003, \mnras, 344, 1250,
  \dodoi{10.1046/j.1365-8711.2003.06912.x}

\bibitem[{{Lagrange} {et~al.}(2019){Lagrange}, {Meunier}, {Rubini}, {Keppler},
  {Galland}, {Chapellier}, {Michel}, {Balona}, {Beust}, {Guillot}, {Grandjean},
  {Borgniet}, {M{\'e}karnia}, {Wilson}, {Kiefer}, {Bonnefoy}, {Lillo-Box},
  {Pantoja}, {Jones}, {Iglesias}, {Rodet}, {Diaz}, {Zapata}, {Abe}, \&
  {Schmider}}]{2019NatAs...3.1135L}
{Lagrange}, A.~M., {Meunier}, N., {Rubini}, P., {et~al.} 2019, Nature
  Astronomy, 3, 1135, \dodoi{10.1038/s41550-019-0857-1}

\bibitem[{{Lallement} {et~al.}(2014){Lallement}, {Vergely}, {Valette},
  {Puspitarini}, {Eyer}, \& {Casagrande}}]{lallement2014}
{Lallement}, R., {Vergely}, J.~L., {Valette}, B., {et~al.} 2014, \aap, 561,
  A91, \dodoi{10.1051/0004-6361/201322032}

\bibitem[{{Lallement} {et~al.}(2018){Lallement}, {Capitanio}, {Ruiz-Dern},
  {Danielski}, {Babusiaux}, {Vergely}, {Elyajouri}, {Arenou}, \&
  {Leclerc}}]{lallement2018}
{Lallement}, R., {Capitanio}, L., {Ruiz-Dern}, L., {et~al.} 2018, \aap, 616,
  A132, \dodoi{10.1051/0004-6361/201832832}

\bibitem[{{Lecavelier des Etangs} {et~al.}(1996){Lecavelier des Etangs},
  {Scholl}, {Roques}, {Sicardy}, \& {Vidal-Madjar}}]{lecavelier1996}
{Lecavelier des Etangs}, A., {Scholl}, H., {Roques}, F., {Sicardy}, B., \&
  {Vidal-Madjar}, A. 1996, \icarus, 123, 168, \dodoi{10.1006/icar.1996.0147}

\bibitem[{{Leconte} \& {Chabrier}(2012)}]{leconte2012}
{Leconte}, J., \& {Chabrier}, G. 2012, \aap, 540, A20,
  \dodoi{10.1051/0004-6361/201117595}

\bibitem[{{Lightkurve Collaboration} {et~al.}(2018){Lightkurve Collaboration},
  {Cardoso}, {Hedges}, {Gully-Santiago}, {Saunders}, {Cody}, {Barclay}, {Hall},
  {Sagear}, {Turtelboom}, {Zhang}, {Tzanidakis}, {Mighell}, {Coughlin}, {Bell},
  {Berta-Thompson}, {Williams}, {Dotson}, \& {Barentsen}}]{lightkurve2018}
{Lightkurve Collaboration}, {Cardoso}, J. V. d.~M., {Hedges}, C., {et~al.}
  2018, {Lightkurve: Kepler and TESS time series analysis in Python}.
\newblock \doeprint{1812.013}

\bibitem[{{Lindegren} {et~al.}(2021{\natexlab{a}}){Lindegren}, {Bastian},
  {Biermann}, {Bombrun}, {de Torres}, {Gerlach}, {Geyer}, {Hern{\'a}ndez},
  {Hilger}, {Hobbs}, {Klioner}, {Lammers}, {McMillan}, {Ramos-Lerate},
  {Steidelm{\"u}ller}, {Stephenson}, \& {van Leeuwen}}]{lindegren2021a}
{Lindegren}, L., {Bastian}, U., {Biermann}, M., {et~al.} 2021{\natexlab{a}},
  \aap, 649, A4, \dodoi{10.1051/0004-6361/202039653}

\bibitem[{{Lindegren} {et~al.}(2021{\natexlab{b}}){Lindegren}, {Klioner},
  {Hern{\'a}ndez}, {Bombrun}, {Ramos-Lerate}, {Steidelm{\"u}ller}, {Bastian},
  {Biermann}, {de Torres}, {Gerlach}, {Geyer}, {Hilger}, {Hobbs}, {Lammers},
  {McMillan}, {Stephenson}, {Casta{\~n}eda}, {Davidson}, {Fabricius},
  {Gracia-Abril}, {Portell}, {Rowell}, {Teyssier}, {Torra}, {Bartolom{\'e}},
  {Clotet}, {Garralda}, {Gonz{\'a}lez-Vidal}, {Torra}, {Abbas}, {Altmann},
  {Anglada Varela}, {Balaguer-N{\'u}{\~n}ez}, {Balog}, {Barache}, {Becciani},
  {Bernet}, {Bertone}, {Bianchi}, {Bouquillon}, {Brown}, {Bucciarelli},
  {Busonero}, {Butkevich}, {Buzzi}, {Cancelliere}, {Carlucci}, {Charlot},
  {Cioni}, {Crosta}, {Crowley}, {del Peloso}, {del Pozo}, {Drimmel}, {Esquej},
  {Fienga}, {Fraile}, {Gai}, {Garcia-Reinaldos}, {Guerra}, {Hambly}, {Hauser},
  {Jan{\ss}en}, {Jordan}, {Kostrzewa-Rutkowska}, {Lattanzi}, {Liao}, {Licata},
  {Lister}, {L{\"o}ffler}, {Marchant}, {Masip}, {Mignard}, {Mints}, {Molina},
  {Mora}, {Morbidelli}, {Murphy}, {Pagani}, {Panuzzo}, {Pe{\~n}alosa Esteller},
  {Poggio}, {Re Fiorentin}, {Riva}, {Sagrist{\`a} Sell{\'e}s}, {Sanchez
  Gimenez}, {Sarasso}, {Sciacca}, {Siddiqui}, {Smart}, {Souami}, {Spagna},
  {Steele}, {Taris}, {Utrilla}, {van Reeven}, \& {Vecchiato}}]{lindegren2021b}
{Lindegren}, L., {Klioner}, S.~A., {Hern{\'a}ndez}, J., {et~al.}
  2021{\natexlab{b}}, \aap, 649, A2, \dodoi{10.1051/0004-6361/202039709}

\bibitem[{{Liu} {et~al.}(2016){Liu}, {Yong}, {Asplund}, {Ram{\'\i}rez}, \&
  {Mel{\'e}ndez}}]{liu2016a}
{Liu}, F., {Yong}, D., {Asplund}, M., {Ram{\'\i}rez}, I., \& {Mel{\'e}ndez}, J.
  2016, \mnras, 457, 3934, \dodoi{10.1093/mnras/stw247}

\bibitem[{{Lothringer} {et~al.}(2021){Lothringer}, {Rustamkulov}, {Sing},
  {Gibson}, {Wilson}, \& {Schlaufman}}]{lothringer2021}
{Lothringer}, J.~D., {Rustamkulov}, Z., {Sing}, D.~K., {et~al.} 2021, \apj,
  914, 12, \dodoi{10.3847/1538-4357/abf8a9}

\bibitem[{{Madhusudhan} {et~al.}(2014){Madhusudhan}, {Amin}, \&
  {Kennedy}}]{madhusudhan2014}
{Madhusudhan}, N., {Amin}, M.~A., \& {Kennedy}, G.~M. 2014, \apjl, 794, L12,
  \dodoi{10.1088/2041-8205/794/1/L12}

\bibitem[{{Madhusudhan} {et~al.}(2017){Madhusudhan}, {Bitsch}, {Johansen}, \&
  {Eriksson}}]{madhusudhan2017}
{Madhusudhan}, N., {Bitsch}, B., {Johansen}, A., \& {Eriksson}, L. 2017,
  \mnras, 469, 4102, \dodoi{10.1093/mnras/stx1139}

\bibitem[{{Mainzer} {et~al.}(2011){Mainzer}, {Grav}, {Bauer}, {Masiero},
  {McMillan}, {Cutri}, {Walker}, {Wright}, {Eisenhardt}, {Tholen}, {Spahr},
  {Jedicke}, {Denneau}, {DeBaun}, {Elsbury}, {Gautier}, {Gomillion}, {Hand},
  {Mo}, {Watkins}, {Wilkins}, {Bryngelson}, {Del Pino Molina}, {Desai},
  {G{\'o}mez Camus}, {Hidalgo}, {Konstantopoulos}, {Larsen}, {Maleszewski},
  {Malkan}, {Mauduit}, {Mullan}, {Olszewski}, {Pforr}, {Saro}, {Scotti}, \&
  {Wasserman}}]{mainzer2011}
{Mainzer}, A., {Grav}, T., {Bauer}, J., {et~al.} 2011, \apj, 743, 156,
  \dodoi{10.1088/0004-637X/743/2/156}

\bibitem[{{McKinney}(2010)}]{mckinney2010}
{McKinney}, W. 2010, in {P}roceedings of the 9th {P}ython in {S}cience
  {C}onference, ed. {S}t\'efan van~der {W}alt \& {J}arrod {M}illman, 56 -- 61,
  \dodoi{10.25080/Majora-92bf1922-00a}

\bibitem[{{M{\'e}karnia} {et~al.}(2017){M{\'e}karnia}, {Chapellier}, {Guillot},
  {Abe}, {Agabi}, {De Pra}, {Schmider}, {Zwintz}, {Stevenson}, {Wang},
  {Lagrange}, {Bigot}, {Crouzet}, {Fante{\"\i}-Caujolle}, {Christille}, \&
  {Kalas}}]{mekarnia2017}
{M{\'e}karnia}, D., {Chapellier}, E., {Guillot}, T., {et~al.} 2017, \aap, 608,
  L6, \dodoi{10.1051/0004-6361/201732121}

\bibitem[{{Miret-Roig} {et~al.}(2020){Miret-Roig}, {Galli}, {Brandner}, {Bouy},
  {Barrado}, {Olivares}, {Antoja}, {Romero-G{\'o}mez}, {Figueras}, \&
  {Lillo-Box}}]{mireg2020}
{Miret-Roig}, N., {Galli}, P.~A.~B., {Brandner}, W., {et~al.} 2020, \aap, 642,
  A179, \dodoi{10.1051/0004-6361/202038765}

\bibitem[{{Mordasini} {et~al.}(2014){Mordasini}, {Klahr}, {Alibert}, {Miller},
  \& {Henning}}]{mordasini2014}
{Mordasini}, C., {Klahr}, H., {Alibert}, Y., {Miller}, N., \& {Henning}, T.
  2014, \aap, 566, A141, \dodoi{10.1051/0004-6361/201321479}

\bibitem[{{Mordasini} {et~al.}(2016){Mordasini}, {van Boekel}, {Molli{\`e}re},
  {Henning}, \& {Benneke}}]{mordasini2016}
{Mordasini}, C., {van Boekel}, R., {Molli{\`e}re}, P., {Henning}, T., \&
  {Benneke}, B. 2016, \apj, 832, 41, \dodoi{10.3847/0004-637X/832/1/41}

\bibitem[{{Morton}(2015)}]{morton2015}
{Morton}, T.~D. 2015, {isochrones: Stellar model grid package}.
\newblock \doeprint{1503.010}

\bibitem[{{Nelson} {et~al.}(2021){Nelson}, {Ting}, {Hawkins}, {Ji}, {Kamdar},
  \& {El-Badry}}]{nelson2021}
{Nelson}, T., {Ting}, Y.-S., {Hawkins}, K., {et~al.} 2021, \apj, 921, 118,
  \dodoi{10.3847/1538-4357/ac14be}

\bibitem[{{Ness} {et~al.}(2018){Ness}, {Rix}, {Hogg}, {Casey}, {Holtzman},
  {Fouesneau}, {Zasowski}, {Geisler}, {Shetrone}, {Minniti}, {Frinchaboy}, \&
  {Roman-Lopes}}]{ness2018}
{Ness}, M., {Rix}, H.~W., {Hogg}, D.~W., {et~al.} 2018, \apj, 853, 198,
  \dodoi{10.3847/1538-4357/aa9d8e}

\bibitem[{{Notsu} {et~al.}(2020){Notsu}, {Eistrup}, {Walsh}, \&
  {Nomura}}]{notsu2020}
{Notsu}, S., {Eistrup}, C., {Walsh}, C., \& {Nomura}, H. 2020, \mnras, 499,
  2229, \dodoi{10.1093/mnras/staa2944}

\bibitem[{{{\"O}berg} {et~al.}(2011){{\"O}berg}, {Murray-Clay}, \&
  {Bergin}}]{oberg2011}
{{\"O}berg}, K.~I., {Murray-Clay}, R., \& {Bergin}, E.~A. 2011, \apjl, 743,
  L16, \dodoi{10.1088/2041-8205/743/1/L16}

\bibitem[{{Ochsenbein} {et~al.}(2000){Ochsenbein}, {Bauer}, \&
  {Marcout}}]{ochsenbein2000}
{Ochsenbein}, F., {Bauer}, P., \& {Marcout}, J. 2000, \aaps, 143, 23,
  \dodoi{10.1051/aas:2000169}

\bibitem[{{pandas Development Team}(2020)}]{reback2020}
{pandas Development Team}. 2020, pandas-dev/pandas: Pandas, latest,  Zenodo,
  \dodoi{10.5281/zenodo.3509134}

\bibitem[{{Paxton} {et~al.}(2011){Paxton}, {Bildsten}, {Dotter}, {Herwig},
  {Lesaffre}, \& {Timmes}}]{paxton2011}
{Paxton}, B., {Bildsten}, L., {Dotter}, A., {et~al.} 2011, \apjs, 192, 3,
  \dodoi{10.1088/0067-0049/192/1/3}

\bibitem[{{Paxton} {et~al.}(2013){Paxton}, {Cantiello}, {Arras}, {Bildsten},
  {Brown}, {Dotter}, {Mankovich}, {Montgomery}, {Stello}, {Timmes}, \&
  {Townsend}}]{paxton2013}
{Paxton}, B., {Cantiello}, M., {Arras}, P., {et~al.} 2013, \apjs, 208, 4,
  \dodoi{10.1088/0067-0049/208/1/4}

\bibitem[{{Paxton} {et~al.}(2015){Paxton}, {Marchant}, {Schwab}, {Bauer},
  {Bildsten}, {Cantiello}, {Dessart}, {Farmer}, {Hu}, {Langer}, {Townsend},
  {Townsley}, \& {Timmes}}]{paxton2015}
{Paxton}, B., {Marchant}, P., {Schwab}, J., {et~al.} 2015, \apjs, 220, 15,
  \dodoi{10.1088/0067-0049/220/1/15}

\bibitem[{{Paxton} {et~al.}(2018){Paxton}, {Schwab}, {Bauer}, {Bildsten},
  {Blinnikov}, {Duffell}, {Farmer}, {Goldberg}, {Marchant}, {Sorokina},
  {Thoul}, {Townsend}, \& {Timmes}}]{paxton2018}
{Paxton}, B., {Schwab}, J., {Bauer}, E.~B., {et~al.} 2018, \apjs, 234, 34,
  \dodoi{10.3847/1538-4365/aaa5a8}

\bibitem[{{Paxton} {et~al.}(2019){Paxton}, {Smolec}, {Schwab}, {Gautschy},
  {Bildsten}, {Cantiello}, {Dotter}, {Farmer}, {Goldberg}, {Jermyn}, {Kanbur},
  {Marchant}, {Thoul}, {Townsend}, {Wolf}, {Zhang}, \& {Timmes}}]{paxton2019}
{Paxton}, B., {Smolec}, R., {Schwab}, J., {et~al.} 2019, \apjs, 243, 10,
  \dodoi{10.3847/1538-4365/ab2241}

\bibitem[{{Piso} {et~al.}(2015){Piso}, {{\"O}berg}, {Birnstiel}, \&
  {Murray-Clay}}]{piso2015}
{Piso}, A.-M.~A., {{\"O}berg}, K.~I., {Birnstiel}, T., \& {Murray-Clay}, R.~A.
  2015, \apj, 815, 109, \dodoi{10.1088/0004-637X/815/2/109}

\bibitem[{{Poovelil} {et~al.}(2020){Poovelil}, {Zasowski}, {Hasselquist},
  {Seth}, {Donor}, {Beaton}, {Cunha}, {Frinchaboy},
  {Garc{\'\i}a-Hern{\'a}ndez}, {Hawkins}, {Kratter}, {Lane}, \&
  {Nitschelm}}]{pooveli2020}
{Poovelil}, V.~J., {Zasowski}, G., {Hasselquist}, S., {et~al.} 2020, \apj, 903,
  55, \dodoi{10.3847/1538-4357/abb93e}

\bibitem[{{Ram{\'\i}rez} {et~al.}(2014){Ram{\'\i}rez}, {Mel{\'e}ndez}, {Bean},
  {Asplund}, {Bedell}, {Monroe}, {Casagrande}, {Schirbel}, {Dreizler}, {Teske},
  {Tucci Maia}, {Alves-Brito}, \& {Baumann}}]{ramirez2014}
{Ram{\'\i}rez}, I., {Mel{\'e}ndez}, J., {Bean}, J., {et~al.} 2014, \aap, 572,
  A48, \dodoi{10.1051/0004-6361/201424244}

\bibitem[{{Reggiani} {et~al.}(2022){Reggiani}, {Schlaufman}, {Healy},
  {Lothringer}, \& {Sing}}]{reggiani2022}
{Reggiani}, H., {Schlaufman}, K.~C., {Healy}, B.~F., {Lothringer}, J.~D., \&
  {Sing}, D.~K. 2022, \aj, 163, 159, \dodoi{10.3847/1538-3881/ac4d9f}

\bibitem[{{Reggiani} {et~al.}(2019){Reggiani}, {Amarsi}, {Lind}, {Barklem},
  {Zatsarinny}, {Bartschat}, {Fursa}, {Bray}, {Spina}, \&
  {Mel{\'e}ndez}}]{reggiani2019}
{Reggiani}, H., {Amarsi}, A.~M., {Lind}, K., {et~al.} 2019, \aap, 627, A177,
  \dodoi{10.1051/0004-6361/201935156}

\bibitem[{{Riello} {et~al.}(2018){Riello}, {De Angeli}, {Evans}, {Busso},
  {Hambly}, {Davidson}, {Burgess}, {Montegriffo}, {Osborne}, {Kewley},
  {Carrasco}, {Fabricius}, {Jordi}, {Cacciari}, {van Leeuwen}, \&
  {Holland}}]{riello2018}
{Riello}, M., {De Angeli}, F., {Evans}, D.~W., {et~al.} 2018, \aap, 616, A3,
  \dodoi{10.1051/0004-6361/201832712}

\bibitem[{{Saffe} {et~al.}(2021){Saffe}, {Miquelarena}, {Alacoria}, {Flores},
  {Jaque Arancibia}, {Calvo}, {Mart{\'\i}n Girardi}, {Grosso}, \&
  {Collado}}]{saffe2021}
{Saffe}, C., {Miquelarena}, P., {Alacoria}, J., {et~al.} 2021, \aap, 647, A49,
  \dodoi{10.1051/0004-6361/202040132}

\bibitem[{{Shectman} \& {Johns}(2003)}]{shectman2003}
{Shectman}, S.~A., \& {Johns}, M. 2003, in Society of Photo-Optical
  Instrumentation Engineers (SPIE) Conference Series, Vol. 4837, \procspie, ed.
  J.~M. {Oschmann} \& L.~M. {Stepp}, 910--918, \dodoi{10.1117/12.457909}

\bibitem[{{Skrutskie} {et~al.}(2006){Skrutskie}, {Cutri}, {Stiening},
  {Weinberg}, {Schneider}, {Carpenter}, {Beichman}, {Capps}, {Chester},
  {Elias}, {Huchra}, {Liebert}, {Lonsdale}, {Monet}, {Price}, {Seitzer},
  {Jarrett}, {Kirkpatrick}, {Gizis}, {Howard}, {Evans}, {Fowler}, {Fullmer},
  {Hurt}, {Light}, {Kopan}, {Marsh}, {McCallon}, {Tam}, {Van Dyk}, \&
  {Wheelock}}]{skrutskie2006}
{Skrutskie}, M.~F., {Cutri}, R.~M., {Stiening}, R., {et~al.} 2006, \aj, 131,
  1163, \dodoi{10.1086/498708}

\bibitem[{{Sneden}(1973)}]{sneden1973}
{Sneden}, C.~A. 1973, PhD thesis, THE UNIVERSITY OF TEXAS AT AUSTIN.

\bibitem[{{Souto} {et~al.}(2022){Souto}, {Cunha}, {Smith}, {Allende Prieto},
  {Covey}, {Garc{\'\i}a-Hern{\'a}ndez}, {Holtzman}, {J{\"o}nsson}, {Mahadevan},
  {Majewski}, {Masseron}, {Pinsonneault}, {Schneider}, {Shetrone}, {Stassun},
  {Terrien}, {Zamora}, {Stringfellow}, {Lane}, {Nitschelm}, \&
  {Rojas-Ayala}}]{souto2022}
{Souto}, D., {Cunha}, K., {Smith}, V.~V., {et~al.} 2022, \apj, 927, 123,
  \dodoi{10.3847/1538-4357/ac4891}

\bibitem[{{Thiabaud} {et~al.}(2015){Thiabaud}, {Marboeuf}, {Alibert}, {Leya},
  \& {Mezger}}]{thiabaud2015}
{Thiabaud}, A., {Marboeuf}, U., {Alibert}, Y., {Leya}, I., \& {Mezger}, K.
  2015, \aap, 574, A138, \dodoi{10.1051/0004-6361/201424868}

\bibitem[{{Ting} {et~al.}(2012){Ting}, {De Silva}, {Freeman}, \&
  {Parker}}]{ting2012}
{Ting}, Y.-S., {De Silva}, G.~M., {Freeman}, K.~C., \& {Parker}, S.~J. 2012,
  \mnras, 427, 882, \dodoi{10.1111/j.1365-2966.2012.22028.x}

\bibitem[{{Tody}(1986)}]{tody1986}
{Tody}, D. 1986, Society of Photo-Optical Instrumentation Engineers (SPIE)
  Conference Series, Vol. 627, {The IRAF Data Reduction and Analysis System},
  ed. D.~L. {Crawford}, 733, \dodoi{10.1117/12.968154}

\bibitem[{{Tody}(1993)}]{tody1993}
---. 1993, Astronomical Society of the Pacific Conference Series, Vol.~52,
  {IRAF in the Nineties}, ed. R.~J. {Hanisch}, R.~J.~V. {Brissenden}, \&
  J.~{Barnes}, 173

\bibitem[{{Torra} {et~al.}(2021){Torra}, {Casta{\~n}eda}, {Fabricius},
  {Lindegren}, {Clotet}, {Gonz{\'a}lez-Vidal}, {Bartolom{\'e}}, {Bastian},
  {Bernet}, {Biermann}, {Garralda}, {G{\'u}rpide}, {Lammers}, {Portell}, \&
  {Torra}}]{torra2021}
{Torra}, F., {Casta{\~n}eda}, J., {Fabricius}, C., {et~al.} 2021, \aap, 649,
  A10, \dodoi{10.1051/0004-6361/202039637}

\bibitem[{{Tremblin} {et~al.}(2016){Tremblin}, {Amundsen}, {Chabrier},
  {Baraffe}, {Drummond}, {Hinkley}, {Mourier}, \&
  {Venot}}]{2016ApJ...817L..19T}
{Tremblin}, P., {Amundsen}, D.~S., {Chabrier}, G., {et~al.} 2016, \apjl, 817,
  L19, \dodoi{10.3847/2041-8205/817/2/L19}

\bibitem[{{Tremblin} {et~al.}(2015){Tremblin}, {Amundsen}, {Mourier},
  {Baraffe}, {Chabrier}, {Drummond}, {Homeier}, \&
  {Venot}}]{2015ApJ...804L..17T}
{Tremblin}, P., {Amundsen}, D.~S., {Mourier}, P., {et~al.} 2015, \apjl, 804,
  L17, \dodoi{10.1088/2041-8205/804/1/L17}

\bibitem[{{Tremblin} {et~al.}(2017){Tremblin}, {Chabrier}, {Mayne}, {Amundsen},
  {Baraffe}, {Debras}, {Drummond}, {Manners}, \&
  {Fromang}}]{2017ApJ...841...30T}
{Tremblin}, P., {Chabrier}, G., {Mayne}, N.~J., {et~al.} 2017, \apj, 841, 30,
  \dodoi{10.3847/1538-4357/aa6e57}

\bibitem[{{Virtanen} {et~al.}(2020){Virtanen}, {Gommers}, {Oliphant},
  {Haberland}, {Reddy}, {Cournapeau}, {Burovski}, {Peterson}, {Weckesser},
  {Bright}, {van der Walt}, {Brett}, {Wilson}, {Millman}, {Mayorov}, {Nelson},
  {Jones}, {Kern}, {Larson}, {Carey}, {Polat}, {Feng}, {Moore}, {VanderPlas},
  {Laxalde}, {Perktold}, {Cimrman}, {Henriksen}, {Quintero}, {Harris},
  {Archibald}, {Ribeiro}, {Pedregosa}, {van Mulbregt}, \& {SciPy 1. 0
  Contributors}}]{virtanen2020}
{Virtanen}, P., {Gommers}, R., {Oliphant}, T.~E., {et~al.} 2020, Nature
  Methods, 17, 261, \dodoi{10.1038/s41592-019-0686-2}

\bibitem[{{Wenger} {et~al.}(2000){Wenger}, {Ochsenbein}, {Egret}, {Dubois},
  {Bonnarel}, {Borde}, {Genova}, {Jasniewicz}, {Lalo{\"e}}, {Lesteven}, \&
  {Monier}}]{wenger2000}
{Wenger}, M., {Ochsenbein}, F., {Egret}, D., {et~al.} 2000, \aaps, 143, 9,
  \dodoi{10.1051/aas:2000332}

\bibitem[{{Wilson} {et~al.}(2017){Wilson}, {Lecavelier des Etangs},
  {Vidal-Madjar}, {Bourrier}, {H{\'e}brard}, {Kiefer}, {Beust}, {Ferlet}, \&
  {Lagrange}}]{wilson2017}
{Wilson}, P.~A., {Lecavelier des Etangs}, A., {Vidal-Madjar}, A., {et~al.}
  2017, \aap, 599, A75, \dodoi{10.1051/0004-6361/201629293}

\bibitem[{{Wright} {et~al.}(2010){Wright}, {Eisenhardt}, {Mainzer}, {Ressler},
  {Cutri}, {Jarrett}, {Kirkpatrick}, {Padgett}, {McMillan}, {Skrutskie},
  {Stanford}, {Cohen}, {Walker}, {Mather}, {Leisawitz}, {Gautier}, {McLean},
  {Benford}, {Lonsdale}, {Blain}, {Mendez}, {Irace}, {Duval}, {Liu}, {Royer},
  {Heinrichsen}, {Howard}, {Shannon}, {Kendall}, {Walsh}, {Larsen}, {Cardon},
  {Schick}, {Schwalm}, {Abid}, {Fabinsky}, {Naes}, \& {Tsai}}]{wright2010}
{Wright}, E.~L., {Eisenhardt}, P. R.~M., {Mainzer}, A.~K., {et~al.} 2010, \aj,
  140, 1868, \dodoi{10.1088/0004-6256/140/6/1868}

\bibitem[{{Yana Galarza} {et~al.}(2019){Yana Galarza}, {Mel{\'e}ndez},
  {Lorenzo-Oliveira}, {Valio}, {Reggiani}, {Carlos}, {Ponte}, {Spina},
  {Haywood}, \& {Gandolfi}}]{galarza2019}
{Yana Galarza}, J., {Mel{\'e}ndez}, J., {Lorenzo-Oliveira}, D., {et~al.} 2019,
  \mnras, 490, L86, \dodoi{10.1093/mnrasl/slz153}

\bibitem[{{Z{\'u}{\~n}iga-Fern{\'a}ndez}
  {et~al.}(2021){Z{\'u}{\~n}iga-Fern{\'a}ndez}, {Bayo}, {Elliott}, {Zamora},
  {Corval{\'a}n}, {Haubois}, {Corral-Santana}, {Olofsson}, {Hu{\'e}lamo},
  {Sterzik}, {Torres}, {Quast}, \& {Melo}}]{zuniga2021}
{Z{\'u}{\~n}iga-Fern{\'a}ndez}, S., {Bayo}, A., {Elliott}, P., {et~al.} 2021,
  \aap, 645, A30, \dodoi{10.1051/0004-6361/202037830}

\end{thebibliography}
\bibliographystyle{aasjournal}

\end{document}